\documentclass[aps,prx,twocolumn,english,superscriptaddress]{revtex4-2}

\usepackage{natbib}
\usepackage{amsfonts, amsmath, amssymb, mathrsfs, mathtools, amsthm, bbm, bm}
\usepackage{blkarray}
\usepackage{braket}
\usepackage{float}
\usepackage{afterpage}
\usepackage[T1]{fontenc}
\usepackage[utf8]{inputenc}
\usepackage{lmodern}
\usepackage[version=4]{mhchem}
\usepackage[labelfont=bf, font=small]{caption}
\usepackage[font=footnotesize]{subcaption}
\usepackage{tikz}
\usetikzlibrary {shapes.geometric, calc}
\usetikzlibrary{decorations.pathreplacing} 
\usetikzlibrary{arrows}
\usepackage{quantikz}
\usepackage{booktabs, tabularx}

\usepackage{pifont}

\usepackage{xcolor,colortbl}

\DeclareMathOperator{\Tr}{Tr}

\newcolumntype{Y}{>{\raggedleft\arraybackslash}X}

\definecolor{N2_plasma0}{rgb}{0.050383, 0.029803, 0.527975}
\definecolor{N2_plasma1}{rgb}{0.417642, 0.000564, 0.658390}
\definecolor{N2_plasma2}{rgb}{0.69284,  0.165141, 0.564522}
\definecolor{N2_plasma3}{rgb}{0.881443, 0.392529, 0.383229}
\definecolor{N2_plasma4}{rgb}{0.98826,  0.652325, 0.211364}

\definecolor{selection}{rgb}{0.417642, 0.000564, 0.658390}

\usepackage[bookmarks=false]{hyperref}

\tikzset{
    my hex/.style={regular polygon, regular polygon sides=6, draw, inner sep=0pt, outer sep=0pt, minimum size=1cm},
    my hex select/.style={regular polygon, regular polygon sides=6, draw, inner sep=0pt, outer sep=0pt, minimum size=1cm, fill=selection!20!white},
    my tri select up/.style={regular polygon, regular polygon sides=3, draw, inner sep=0pt, outer sep=0pt, minimum size=2*0.866cm, fill=selection!20!white},
    my tri select down/.style={regular polygon, regular polygon sides=3, draw, inner sep=0pt, outer sep=0pt, minimum size=2*0.866cm, fill=selection!20!white, rotate=180},
    my circ/.style={draw, circle, fill=black, inner sep=0pt, minimum size=0.9mm},
    my circ select/.style={draw, circle, fill=selection, inner sep=0pt, minimum size=1.7mm}
}

\begin{document}
\title{Simulating the Antiferromagnetic Heisenberg Model on a Spin-Frustrated Kagome Lattice with the Contextual Subspace Variational Quantum Eigensolver}

\author{Tim Weaving}
\affiliation{Centre for Computational Science, Department of Chemistry, University College London, WC1H 0AJ, United Kingdom}
\author{Alexis Ralli}
\affiliation{Centre for Computational Science, Department of Chemistry, University College London, WC1H 0AJ, United Kingdom}
\author{Vinul Wimalaweera}
\affiliation{London Centre for Nanotechnology, University College London, WC1H 0AH, United Kingdom}
\author{Peter J. Love}
\affiliation{Department of Physics and Astronomy, Tufts University, Medford, MA 02155, USA}
\affiliation{Computational Science Initiative, Brookhaven National Laboratory, Upton, NY 11973, USA}
\author{Peter V. Coveney}
\affiliation{Centre for Computational Science, Department of Chemistry, University College London, WC1H 0AJ, United Kingdom}
\affiliation{Advanced Research Computing Centre, University College London, WC1H 0AJ, United Kingdom}
\affiliation{Informatics Institute, University of Amsterdam, Amsterdam, 1098 XH, Netherlands}

\date{\today}

\begin{abstract}
In this work we investigate the ground state properties of a candidate quantum spin liquid using a superconducting Noisy Intermediate-Scale Quantum (NISQ) device. Specifically, we study the antiferromagnetic Heisenberg model on a Kagome lattice, a geometrically frustrated structure that gives rise to a highly degenerate energy spectrum. To successfully simulate this system, we employ a qubit reduction strategy leveraging the Contextual Subspace methodology, significantly reducing the problem size prior to execution on the quantum device. We improve the quality of these subspaces by using the wavefunctions obtained from low bond dimension Density Matrix Renormalization Group (DMRG) calculations to bias the subspace stabilizers through a symplectic approximate symmetry generator extraction algorithm. Reducing the Hamiltonian size allows us to implement tiled circuit ensembles and deploy the Variational Quantum Eigensolver (VQE) to estimate the ground state energy. We adopt a hybrid quantum error mitigation strategy combining Readout Error Mitigation (REM), Symmetry Verification (SV) and Zero Noise Extrapolation (ZNE). This approach yields high-accuracy energy estimates, achieving error rates on the order of 0.01\% and thus demonstrating the potential of near-term quantum devices for probing frustrated quantum materials.
\end{abstract}

\maketitle

\section{Introduction}


The Heisenberg model serves as a foundational framework in quantum many-body physics and condensed matter theory for probing the collective behavior of interacting spin systems. Its versatility and simplicity make it an ideal platform for exploring a range of emergent quantum phenomena, including magnetic ordering, critical points and quantum phase transitions. In particular, the model encapsulates exchange interactions between spin-$\frac{1}{2}$ particles localized on a lattice and captures essential features of magnetic materials.

An $N$-site Heisenberg Hamiltonian with interaction graph specified by edge set $\mathcal{E} \subset \mathbb{Z}_N^{\times2}$, coupling strengths $J_x, J_y, J_z \in \mathbb{R}$ along each spin component and external field with strength $h \in \mathbb{R}$ is defined as
\begin{equation}\label{H}
    \small
    H = - \sum_{(k,\ell) \in \mathcal{E}} \Big[J_x \sigma_x^{(k)} \sigma_x^{(\ell)} + J_y \sigma_y^{(k)} \sigma_y^{(\ell)} + J_z \sigma_z^{(k)} \sigma_z^{(\ell)}\Big] - h \sum_{n=0}^{N-1} \sigma_z^{(n)},
\end{equation}
where $\sigma_x, \sigma_y, \sigma_z$ are the Pauli spin operators. 
The anisotropic case of $J_x \neq J_y \neq J_z$ is referred to as the $XYZ$ model, but there are restricted configurations that are of interest. The one we consider here is the $XXX$ model, characterised by $J_x = J_y = J_z = J$. For $J>0$ the $XXX$ model exhibits ferromagnetic behaviour, whereas for $J<0$ it is antiferromagnetic. Note that, by turning off couplings in certain directions, we may reduce Heisenberg to the Ising model, namely with $J_x=J_y=0$ we get the classical statistical model and $J_z=J_y=0$ yields a rotated transverse-field quantum system.

\begin{figure}[b]
    \centering
    \resizebox{0.85\linewidth}{!}{
    \begin{tikzpicture}[thick]
\draw (-5/2,{sqrt(3)}) -- (-11/4,{sqrt(3)*(3/4)});
\draw (-5/2,{sqrt(3)}) -- (-11/4,{sqrt(3)*(5/4)});
\draw (-5/2,{2*sqrt(3)}) -- (-11/4,{sqrt(3)*(7/4)});
\draw (-5/2,{2*sqrt(3)}) -- (-11/4,{sqrt(3)*(9/4)});

\draw (+7/2,{sqrt(3)}) -- (+15/4,{sqrt(3)*(3/4)});
\draw (+7/2,{sqrt(3)}) -- (+15/4,{sqrt(3)*(5/4)});
\draw (+7/2,{2*sqrt(3)}) -- (+15/4,{sqrt(3)*(7/4)});
\draw (+7/2,{2*sqrt(3)}) -- (+15/4,{sqrt(3)*(9/4)});

\draw (-9/4,{sqrt(3)/4}) -- (+13/4,{sqrt(3)/4});
\draw (-9/4,{11*sqrt(3)/4}) -- (+13/4,{11*sqrt(3)/4});

  \foreach \n[count=\k] in {7,6,7,6,7}{
    \foreach \m in {1,...,\n}{
      \node(h)[my hex] at (-\n/2+\m,{\k*sqrt(3)/2}){};
      \foreach \t in {1,...,6} \node[my circ] at ($(h)+({(\t-1)*60}:.5)$){};
    }
  }
  \node[my tri select up] at (-7/2+4,{6*sqrt(3)/4}){};
  \node[my tri select down] at (-7/2+4,{6*sqrt(3)/4}){};
  
  \node(h)[my hex select] at (-7/2+4,{3*sqrt(3)/2}){};
  \foreach \t in {1,...,6} \node[my circ select] at ($(h)+({(\t-1)*60}:.5)$){};
  \foreach \t in {1,...,6} \node[my circ select] at ($(h)+{sqrt(3)}*({(\t-1/2)*60}:.5)$){};
  
\end{tikzpicture}
    }
    \caption{The Kagome lattice structure, consisting of hexagonal cells with joining triangular elements; the spin-frustration manifests in the latter geometric property. We isolate the central star-shaped lattice consisting of 12 spin-sites for our quantum simulation, highlighted above.}
    \label{fig:kagome_lattice}
\end{figure}
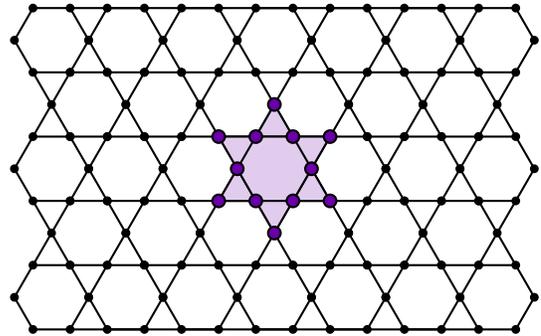

In this work, we investigate a single cell of the Kagome lattice depicted in Figure \ref{fig:kagome_lattice} and defined by the edge-set: 
\begin{equation}
\small
\begin{aligned}
    \mathcal{E} ={} \{ 
    & (0, 1), (1, 2), (2, 3),  (3, 4),  (4, 5),  (0, 5), \\
    & (0, 6), (1, 6), (1, 7),  (2, 7),  (2, 8),  (3, 8), \\
    & (3, 9), (4, 9), (4, 10), (5, 10), (5, 11), (0, 11)\}.
\end{aligned}
\end{equation}

Its geometrical structure gives rise to some exotic physics. Comprised of a two-dimensional arrangement of corner-sharing triangles, this is a canonical example of a highly frustrated lattice geometry \cite{balents2010spin}. In the antiferromagnetic regime ($J < 0$), competing exchange interactions prevent simultaneous minimization of all pairwise spin interactions on triangular plaquettes \cite{nikolic2005theory} and can thus form pinwheel valence-bond state \cite{matan2010pinwheel}. This results in ground state degeneracy and a suppression of long-range magnetic order, even at zero temperature. The Kagome lattice is a candidate quantum spin liquid (QSL), a novel state of matter characterized by the presence of long-range quantum entanglement, topological order and fractionalized excitations \cite{han2012fractionalized}.

\begin{figure}[b!]
    \centering
    \includegraphics[width=0.95\linewidth]{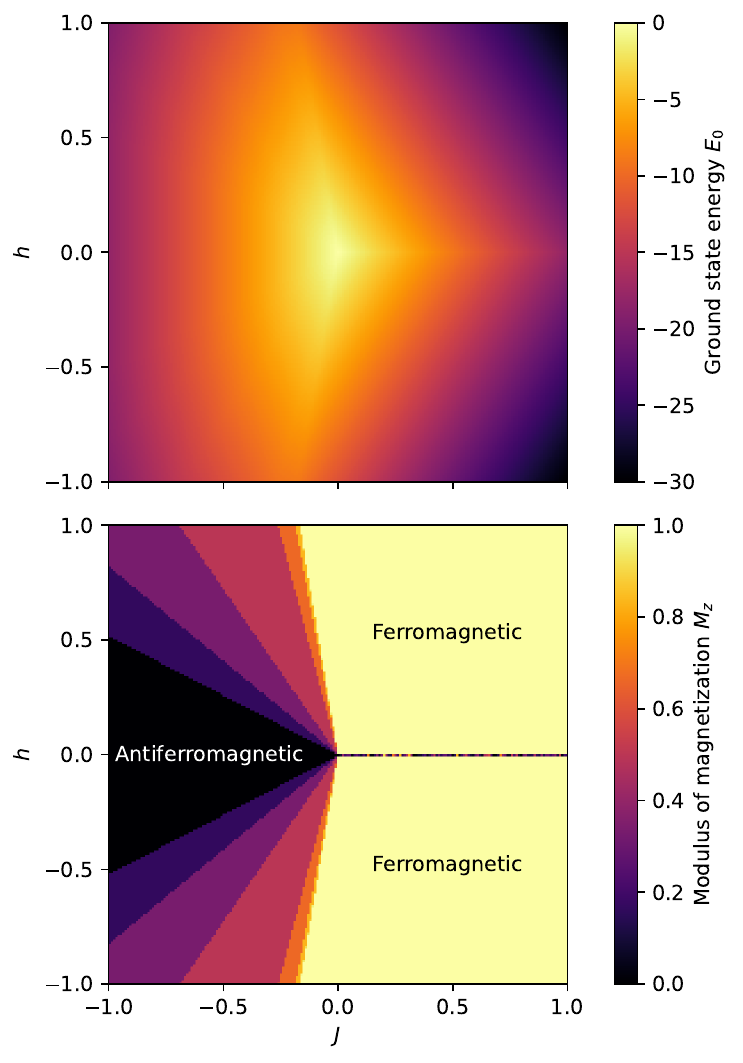}
    \caption{The Heisenberg $XXX$ model ground state energy and modulus of magnetization with varying coupling and field strengths $J$ and $h$ for a single cell of the Kagome lattice, highlighted in Figure \ref{fig:kagome_lattice}. The $E_0$ and $M_z$ values were computed using exact diagonalization of the Hamiltonian.}
    \label{fig:kagome_ground_state}
\end{figure}

In Figure \ref{fig:kagome_ground_state} we present the ground state energy $E_0$ and modulus of magnetization $M_z  = |\sum_{n=0}^{N-1} \braket{\sigma_z^{(n)}}|/N$ for varying $J \in [-1,1]$ and $h \in [0,1]$. We expect a ferromagnetic system to yield $M_z=1$, where all spins are aligned up or down. By contrast, in the ideal antiferromagnetic setting all spins cancel to give $M_z = 0$. However, spin frustration can disrupt perfect ordering and therefore yield a paramagnet with nonzero modulus of magnetization. In the presence of a magnetic field, we see a quantization of the magnetization levels for $J<0$ in Figure \ref{fig:kagome_ground_state}, indicated by discrete bands in the $M_z$ plot.

For the purposes of this work we focus on the field-free Heisenberg $XXX$ model with $J = -1$ and $h=0$, its Hamiltonian $ H = \sum_{(k,\ell) \in \mathcal{E}} \bm{\sigma}^{(k)} \cdot \bm{\sigma}^{(\ell)}$ describing the regime in which antiferromagnetic behaviour is encountered. This Hamiltonian has a doubly degenerate ground state with eigenvalue $E_0 = -18$, which we take as our target energy. Our goal is to prepare the ground state of this system using the Variational Quantum Eigensolver (VQE) \cite{peruzzo2014variational} deployed on the \texttt{ibmq\_guadalupe} Falcon r4P chip, with its qubit layout shown in Figure \ref{fig:guadalupe_topology}. 

Several works have studied the Kagome lattice using VQE \cite{bosse2022probing, kattemolle2022variational}; the approach we take here leverages the Contextual Subspace methodology \cite{kirby2021contextual, weaving2023stabilizer, ralli2023unitary} and demonstrates its relevance to condensed matter, where previous works have looked exclusively at chemical applications~\cite{weaving2023benchmarking, weaving2023contextual, liang2023spacepulse, kiser2025contextual, yao2025quantum, bickley2025extending}. While we study ground state properties here, our subspace approach is also compatible with other workloads such as Hamiltonian simulation. In a previous work we studied the time evolution of spin magnetization in a transverse field Ising model on superconducting hardware to showcase a novel error mitigation approach combining Echo Verification and Clifford Data Regression \cite{weaving2025accurately}. While we did not apply subspace methods there, it could be an avenue for future research.


\section{Qubit Subspace Approach}\label{subspace}

To improve simulation results, it is possible to pre-process the Hamiltonian by analyzing its symmetries to identify possible qubit reductions. First we perform Qubit Tapering  \cite{bravyi2017tapering, setia2020reducing} in Section \ref{sec:tapering} to enforce physical $\mathbb{Z}_2$-symmetry and thus preserve the spectrum exactly. In Section \ref{sec:contextual_subspace} we then seek further qubit reductions through the Contextual Subspace approach \cite{kirby2021contextual, weaving2023stabilizer, ralli2023unitary, weaving2023contextual}, which identifies different (near) stabilizer subspaces which are subsequently enforced in a similar way as in Qubit Tapering.

\subsection{Qubit Tapering}\label{sec:tapering}

In the setting of a Pauli Hamiltonian $H = \sum_m h_m \bm{\sigma}_m$ where $\bm{\sigma}$ denotes an $N$-fold tensor product of single-qubit Pauli matrices, i.e. $\bm{\sigma \in }\{I,\sigma_x,\sigma_y,\sigma_z\}^{\otimes N}$, a $\mathbb{Z}_2$-type symmetry is a Pauli operator $S$ such that $[S,\bm{\sigma}_m]=0\;\forall m$. This is a more restricted notion of symmetry than in the general sense, which can mean any observable that commutes with the Hamiltonian, but not necessarily with individual Hamiltonian terms. For a Heisenberg Hamiltonian, defined in Equation \eqref{H}, there is always a $\mathbb{Z}_2$-type symmetry $S_z=\sigma_z^{\otimes N}$ that encodes spin parity. In the absence of an external field (that would otherwise contribute additional Hamiltonian terms $\sum_{n} \sigma_z^{(n)}$), there is a second $\mathbb{Z}_2$-type symmetry $S_x=\sigma_x^{\otimes N}$ relating to the spin-flip symmetry. 

Qubit Tapering \cite{bravyi2017tapering} makes the observation that each $\mathbb{Z}_2$ symmetry equates to the degree of freedom for a single qubit. By the theory of stabilizer codes \cite{gottesman1997stabilizer}, there exists a Clifford rotation mapping each $\mathbb{Z}_2$ symmetry onto a single qubit position, which may subsequently be projected out of the Hamiltonian. In this example, tapering allows us to remove two degrees of freedom corresponding with the $S_z,S_x$ symmetries, reducing the full 12-qubit Hamiltonian $H$ in Equation \eqref{H} to a 10-qubit system $H_{\mathrm{taper}}$ that is isospectral, i.e. its energy spectrum is preserved exactly up to a potential lifting of various degeneracies. 

Explictly, the Clifford rotation
\begin{equation}
    R = e^{-i \frac{\pi}{4} \sigma_y^{(1)}} e^{\frac{\pi}{4} \sigma_z^{(1)} \sigma_x^{\otimes N}} e^{\frac{\pi}{4} \sigma_x^{(0)} \sigma_z^{\otimes N}}
\end{equation}
allows us to map the $\mathbb{Z}_2$-type symmetries onto single-qubit Pauli operators $R \sigma_z^{\otimes N} R^\dag = \sigma_x^{(0)}, R \sigma_x^{\otimes N} R^\dag = \sigma_x^{(1)}$ acting on distinct qubit positions. Since conjugation by a unitary preserves commutation, the rotated hamiltonian $R H R^\dag$ will necessarily consist of identity or $\sigma_x$ operations in the qubit positions $\{0,1\}$ onto which we mapped the symmetry operators. We may therefore drop these qubits, leaving in their place an eigenvalue corresponding with the relevant symmetry sector, and thus defining an effective $(N-2)$-qubit Hamiltonian
\begin{equation}\label{H_taper}
\begin{aligned}
    & H_{\mathrm{taper}} ={} \\
     &- \sum_{\substack{(k,\ell) \in \mathcal{E} \\ k,\ell \notin \{0,1\}}} \Big[J_x \sigma_x^{(k^\prime)} \sigma_x^{(\ell^\prime)} + J_y \sigma_y^{(k^\prime)} \sigma_y^{(\ell^\prime)} + J_z \sigma_z^{(k^\prime)} \sigma_z^{(\ell^\prime)}\Big] \\ 
    &  +  J_x \sum_{(0, \ell) \in \mathcal{E}} (I + \sigma_x^{\otimes N^\prime}) \sigma_x^{(\ell^\prime)}
      -  J_z \sum_{(1, \ell) \in \mathcal{E}} (I + \sigma_z^{\otimes N^\prime}) \sigma_z^{(\ell^\prime)} \\
    &  - J_y \sum_{(1, \ell) \in \mathcal{E}} \Big[\sigma_z^{\otimes \ell^\prime}\sigma_x^{(\ell^\prime)}\sigma_z^{\otimes(N^\prime-\ell^\prime-1)} + \sigma_x^{\otimes \ell^\prime}\sigma_z^{(\ell^\prime)}\sigma_x^{\otimes(N^\prime-\ell^\prime-1)} \Big]
\end{aligned}
\end{equation}
where $k^\prime = k-2$ denotes an index shift. We reiterate that this holds only for the field free case of $h=0$; for non-zero field strength there exists just a single $\mathbb{Z}_2$ symmetry and therefore it is only possible to remove a single qubit in that setting. In the isotropic, antiferromagnetic regime of $J_x=J_y=J_z=-1$, the ground eigenspace for our Kagome system remains doubly degenerate, indicating the presence of some non-$\mathbb{Z}_2$ symmetry that, if found, we may be able to use for quantum error mitigation purposes \cite{weaving2023benchmarking}.

\subsection{Contextual Subspace}\label{sec:contextual_subspace}

In the previous Section \ref{sec:tapering}, it was demonstrated how the presence of symmetry can be leveraged for a reduction in quantum resource, namely through projection onto an effective Hamiltonian over a reduced set of qubits. The mechanism facilitating this reduction identifies $\mathbb{Z}_2$-type symmetry with a qubit's worth degree of freedom via Clifford rotations and single-qubit projections. In Qubit Tapering \cite{bravyi2017tapering}, the exploited symmetries are physical in the sense they encode some real property of the underlying system such a spin-parity or spin-flip symmetry and therefore the reduction is exact, meaning the spectrum is unchanged (up to eigenvalue multiplicity). One might also consider what would happen if the requirement of physicality were lifted and instead we permited near or ``pseudo'' symmetry to be enforced through the projection procedure. This is the approach of the Contextual Subspace framework \cite{kirby2021contextual, weaving2023stabilizer, ralli2023unitary, weaving2023contextual}, which is moreover built on the quantum foundational concept of strong measurement contextuality \cite{peres1990incompatible, mermin1990simple, mermin1993hidden,spekkens2005contextuality, spekkens2007evidence, spekkens2008negativity, Spekkens2016, kirby2019contextuality, kirby2020classical, raussendorf2020phase}.

From a practical point-of-view, the Contextual Subspace approach involves identifying an independent set of commuting Pauli operators $\mathcal{S}$ (referred to as pseudo-symmetries, which generate a stabilizer group) that we wish to enforce over the Hamiltonian and will subsequently stabilize the resulting contextual subspace. This process is a generalization of Qubit Tapering, which may be formulated within the Contextual Subspace framework by taking $\mathcal{S}$ to be the set of physical $\mathbb{Z}_2$-type symmetries of $H$, in this case $\mathcal{S}=\{S_x,S_z\}$. More generally, choosing a set of size $|\mathcal{S}| = K$ for $K\leq N$ allows a reduction of the Hamiltonian by $K$ qubits, with the contextual subspace defined over $N-K$ qubits. However, the quality of the resulting contextual subspace and corresponding energy estimates is highly sensitive to $\mathcal{S}$ and finding such a set is the core challenge of the method. One approach is to devise a method for identifying near $\mathbb{Z}_2$ symmetries; in the following Section \ref{sec:approx} we demonstrate that this may be achieved via an approximate symmetry generator extraction method that utilizes the symplectic representation of Pauli operators. 

Once a set $\mathcal{S}$ has been identified, the correct sector $\bm{\nu} \in \{\pm1\}^{\times K}$ is found by solving a noncontextual hidden-variable model \cite{kirby2020classical, kirby2021contextual, weaving2023stabilizer}, which can be done classically. The projector onto the contextual subspace is then obtained as $\mathbb{P} = \frac{1}{2^K} \prod_{S \in \mathcal{S}} (\mathbbm{1} + \nu_S S)$, which can be rotated onto single-qubit projections via a rotation $R$, similar to the Qubit Tapering example in the previous Section \ref{sec:tapering}. The rotated projector has the form $\mathbb{P}^\prime = R\mathbb{P}R^\dag = \frac{1}{2^K} \bigotimes_{k=0}^{K-1} (I^{(\mathcal{I}_k)} + \nu_k \sigma_{q_k}^{(\mathcal{I}_k)})$ for some set of qubit indices $\mathcal{I}\subset\mathbb{Z}_N$ satisfying $|\mathcal{I}|=K$. Finally, the reduced contextual subspace Hamiltonian is obtained as
\begin{equation}\label{eq:cs_proj}
    H_{\mathrm{CS}} = \Tr_{\mathcal{I}} \big(\mathbb{P}^\prime RHR^\dagger \mathbb{P}^\prime\big)
\end{equation}
which can be constructed efficiently, for example using the \texttt{symmer} Python package \cite{symmer2022}.

\subsubsection{Identifying Approximate $\mathbb{Z}_2$-type Symmetries in the Symplectic Representation}\label{sec:approx}

In the symplectic representation, a Pauli operator or string $\bm{\sigma} = \bigotimes_{n=0}^{N-1} \sigma^{(n)}$ where $\sigma^{(n)} \in \{I,\sigma_x,\sigma_y,\sigma_z\}$ indicates the Pauli matrix acting in each qubit position can be identified with a pair of binary vectors $\bm{x},\bm{z}$ of length $N$ such that
\begin{equation}\label{eq:symp_def}
    x_{n} = \begin{cases}
        1, & \sigma^{(n)} \in \{\sigma_x,\sigma_y\} \\
        0, & \sigma^{(n)} \in \{I,\sigma_z\}
    \end{cases},
    z_{n} = \begin{cases}
        1, & \sigma^{(n)} \in \{\sigma_z,\sigma_y\} \\
        0, & \sigma^{(n)} \in \{I,\sigma_x\}.
    \end{cases}
\end{equation}
In other words, $\bm{x},\bm{z}$ track $\sigma_x,\sigma_z$ factors in the Pauli string $\bm{\sigma}$, i.e. if $x_{n} = z_{n} = 1$ then $\sigma^{(n)}=\sigma_x\sigma_z=\sigma_y$ up to a multiplicative phase. 
Given a linear combination of $M$ Pauli terms we can store the entire operator as a pair of symplectic matrices $\bm{X},\bm{Z}$ of size $M \times N$ where rows store the symplectic vectors $\bm{x},\bm{z}$ encoding the corresponding Pauli term according to Equation \eqref{eq:symp_def}, along with a vector of coefficients $\bm{c}\in\mathbb{R}^M$. These $X$- and $Z$-components may also be concatenated to form a single $M\times2N$ matrix \mbox{$\bm{B} = \bm{X} | \bm{Z}$} with correspondingly concatenated rows \mbox{$\bm{b} = \bm{x} | \bm{z}$}. Let $\varsigma$ be the mapping from the symplectic representation back into the Pauli group, 
\begin{equation}
    \varsigma\bm(\bm{b}) = i^{\bm{x} \cdot \bm{z}} \bigotimes_{n=0}^{N-1} \sigma_x^{x_n} \sigma_z^{z_n} = \bm{\sigma},
\end{equation}
where the phase $i^{\bm{x} \cdot \bm{z}}$ accounts for products $\sigma_x\sigma_z=-i\sigma_y$.

Many operations over the Pauli group reduce to binary logic in the symplectic representation. Of particular importance here are commutation relations, since this is central to extracting symmetry generating sets. Two Pauli strings $\bm{\sigma}_1,\bm{\sigma}_2$ commute if their pairwise non-identity qubit positions differ in an even number of positions; this amounts to counting mismatches between the $X/Z$ components of their symplectic vectors $\bm{b}_1, \bm{b}_2$. Defining the \textit{canonical symplectic form} $\Omega \coloneqq \begin{bmatrix}
    \bm{0} & \mathbbm{1} \\
    -\mathbbm{1} & \bm{0} \\
\end{bmatrix},$
we may define the symplectic inner product
\begin{equation}
    \braket{\bm{b}_1, \bm{b}_2} \coloneqq \bm{b}_1 \Omega \bm{b}_2^{\intercal} = \bm{x}_1 \cdot \bm{z}_2 - \bm{z}_1 \cdot \bm{x}_2
\end{equation}
and then it follows
\begin{equation}
    \braket{\bm{b}_1, \bm{b}_2} = 0 \mod 2 \Leftrightarrow [\bm{\sigma}_1,\bm{\sigma}_2]=0,
\end{equation}
namely the commutation between Pauli operators $\bm{\sigma}_1,\bm{\sigma}_2$ can be assessed via the inner product defined on their underlying symplectic vectors. This extends similarly to full symplectic matrices $\bm{B}_1, \bm{B}_2$ for operators \mbox{$H_1=\sum_{m=0}^{M_1-1} h_{1,m} \bm{\sigma}_{1,m}, \; H_2=\sum_{m=0}^{M_2-1} h_{2,m} \bm{\sigma}_{2,m}$} with $M_1, M_2$ terms respectively, where 
\begin{equation}
    \braket{\bm{B}_1, \bm{B}_2} = \bm{X}_1\bm{Z}_2^{\intercal} - \bm{Z}_1\bm{X}_2^{\intercal}   
\end{equation}
yields a matrix of size $M_1 \times M_2$ in which, taken modulo 2, element $(m,n)$ is zero if $[\bm{\sigma}_{1,m}, \bm{\sigma}_{2,n}]=0$ and one if $\{\bm{\sigma}_{1,m}, \bm{\sigma}_{2,n}\}=0$. The bit-flipped matrix $\bm{1} - \braket{\bm{B}, \bm{B}}$ for the symplectic inner product of an operator with itself yields the adjacency matrix for its compatibility graph.

This symplectic framework provides a neat way of identifying symmetry generating sets and, through a small modification, \textit{approximate} symmetries. Let $\mathcal{G}$ be an independent generating set for the $\mathbb{Z}_2$-type symmetries of an operator $H$ and $\bm{G}$ be its symplectic matrix. Then we must have $\bm{B} \Omega \bm{G}^{\intercal}=\bm{0}$; note, then, that finding $\mathcal{G}$ is equivalent to determining the kernel or null space of the matrix $\bm{B} \Omega = (-\bm{Z})|\bm{X}$, constructed by swapping the $X$- and $Z$- components of $\bm{B}$. This can be achieved efficiently through Guassian elimination applied to the columns of
$\begin{bmatrix}
    \bm{B} \Omega \\ \hline \mathbbm{1}_{2N}
\end{bmatrix},$
which identifies an invertible matrix $\bm{P}$ satisfying
\begin{equation}\label{eq:col_reduction}
    \begin{bmatrix}
    \bm{B} \Omega \\ \hline \mathbbm{1}_{2N}
\end{bmatrix} \bm{P} = \begin{bmatrix}
    \bm{R} \\ \hline \bm{Q}
\end{bmatrix}.
\end{equation}
One may then observe that $\bm{B} \Omega\bm{P} = \bm{R}$ and $\bm{P}=\bm{Q}$ so that $\bm{B} \Omega\bm{Q} = \bm{R}$. Recalling that our goal here is to identify the kernel of $\bm{B} \Omega$, this can now be read off from the right-hand matrix in Equation \eqref{eq:col_reduction}; taking $\bm{G}^{\intercal}$ to be the columns of $\bm{Q}$ such that the corresponding column in $\bm{R}$ is all zero, we satisfy $\bm{B} \Omega \bm{G}^{\intercal}=\bm{0}$ as required. Explicitly, if $\mathcal{I} \subset \mathbb{Z}_{2N}$ is the set of column indices where $\bm{R}$ is a vector of zeros, i.e. $n\in\mathcal{I}$ if $R_{m,n} = 0 \;\forall m \in \mathbb{Z}_M$, then take $\bm{G}^{\intercal} = \begin{bmatrix}
    \bm{Q}_{:,\mathcal{I}_0} | \bm{Q}_{:,\mathcal{I}_1} | \dots | \bm{Q}_{:,\mathcal{I}_{|\mathcal{I}|-1}}
\end{bmatrix}$
and finally, by transposing, we obtain $\bm{G}$, the symplectic representation of the symmetry generating set $\mathcal{G}$.

This provides a way of identifying the $\mathbb{Z}_2$-type symmetries we used in Section \ref{sec:tapering} for Qubit Tapering, without having \textit{a priori} knowledge of what the physical symmetries for a given Hamiltonian system should be. However, with a slight modification it will also allow us to find approximate symmetries. Above, we found that symmetry generator elements are indexed by columns of zero entries in $\bm{R}$ from Equation \eqref{eq:col_reduction}, which arose as a consequence of the equation $\bm{B} \Omega \bm{Q} = \bm{R}$. However, this relation encodes further information that we may leverage to identify not just exact $\mathbb{Z}_2$-type symmetry, but also the operators that are very close to being symmetries.

Define by
\begin{equation}\label{eq:sym_score}
    w_n \coloneqq \frac{||\bm{c} \cdot (\bm{1} - \bm{R}_{:,\,n})||}{||\bm{c}||} = \sqrt{\frac{\sum_{m=0}^{M-1} |c_m|^2(1 - \bm{R}_{m,n})}{\sum_{m=0}^{M-1} |c_m|^2}}
\end{equation}
the normalized weighted term-wise commutation score. The intuition behind this score is that, for each term $\bm{\sigma}_m$ that commutes with the Pauli whose symplectic representation is stored in column $n$ of $\bm{Q}$, recovering the Pauli as $\varsigma(\bm{Q}_{:,\,n})$, the corresponding coefficient $|c_m|^2$ penalizes the score proportionately. An alternative way of writing this would be
\begin{equation}
        w_n = \sqrt{\frac{\sum_{m: [\bm{\sigma}_m, \varsigma(\bm{Q}_{:,\,n})]=0} |c_m|^2}{\sum_{m=0}^{M-1} |c_m|^2}}.
\end{equation}

Note that $w_n \in [0,1]$ and, if $w_n=1$, this means $[\bm{\sigma}_m, \varsigma(\bm{Q}_{:,\,n})]=0 \;\;\forall m \in \{0, \dots, M-1\}$. In other words, the Pauli $\varsigma(\bm{Q}_{:,\,n})$ corresponding with the $n$-th column of $\bm{Q}$ is a true $\mathbb{Z}_2$-type symmetry, equivalent to the $n$-th column of $\bm{R}$ consisting of all zeros. Conversely, if $w_n=0$ then it must be the case $\{\bm{\sigma}_m, \varsigma(\bm{Q}_{:,\,n})\}=0 \;\;\forall m \in \{0, \dots, M-1\}$, i.e. we have found an antisymmetry generating element, corresponding with a column of ones in $\bm{R}$. However, the real benefit of this reformulation of the symmetry generator extraction procedure comes when we consider the non-integer values of $w_n$. Indeed, if $w_n$ is close to $1$, say $1-w_n<\epsilon$ for some threshold $\epsilon>0$, then $\varsigma(\bm{Q}_{:,\,n})$ is close to being a symmetry, but anticommutes with a subset of Pauli terms whose coefficient norm is below $\epsilon$. We may sort the vector of approximate symmetry weights $\bm{w} = (w_n)_{n=0}^{2N-1}$ to find the best set of approximate $\mathbb{Z}_2$-type symmetries for the input operator with symplectic matrix $\bm{B}$.

This mechanism is central to the Contextual Subspace method, since it relies on the identification of a set of stabilizing elements that define the projection into the subspace as in Equation \eqref{eq:cs_proj}. Using the approximate symmetry generating set approach, together with some reference wavefunction, allows one to bias the stabilizer selection and construct higher-quality contextual subspaces. In the following Section \ref{sec:cs_bias} we investigate the quality of subspaces produced via a low bond dimension tensor network approach using the Density Matrix Renormalization Group (DMRG).

\subsubsection{Biasing Contextual Subspace Stabilizer Selection with the Densitry Matrix Renormalization Group}\label{sec:cs_bias}

In the previous Section \ref{sec:approx} we provided an algorithm for identifying approximate $\mathbb{Z}_2$-type symmetries for some input operator. We now demonstrate how this can be used to select the stabilizers $\mathcal{S}$ that define a particular contextual subspace, motivated by an approximate wavefunction solution that biases this selection. Assuming access to a wavefunction $\ket{\psi}$, we may express this state as a wave operator $\Theta$ applied to reference $\ket{\psi_{\mathrm{ref}}}$, so that $\ket{\psi} = \Theta\ket{\psi_{\mathrm{ref}}}$. Then, using the results of the previous section, we may construct a set of $2N$ approximate symmetries denoted $\mathcal{S}(\Theta)$. Let $\hat{\mathcal{S}}(\Theta) \subset \mathcal{S}(\Theta)$ be a commuting subset of size $N$, ordered by decreasing score $w_n$ in Equation \eqref{eq:sym_score} so that for $S_k, S_\ell \in \hat{\mathcal{S}}(\Theta)$ with $k < \ell$ we have $w_k \geq w_\ell$. Finally, for a contextual subspace over $N_{\mathrm{CS}}$ qubits, we need to choose $K=N - N_{\mathrm{CS}}$ independent commuting operators, which generates a stabilizer group and each stabilizer defines a 1D projection onto its respective eigenspace. In the approximate symmetry framework, we may take 
\begin{equation}\label{eq:stab_select}
    \mathcal{S} = \hat{\mathcal{S}}(\Theta) \Bigr|_{n=0}^{N-N_{\mathrm{CS}}-1} = \{S_0, \dots, S_{N-N_{\mathrm{CS}}-1}\}   
\end{equation}
and, recalling the introduction to Section \ref{sec:contextual_subspace}, the overall projection operator is $\mathbb{P} = \frac{1}{2^K} \prod_{S \in \mathcal{S}} (\mathbbm{1} + \nu_S S)$ for a sector $\bm{\nu}$ that is optimized via a noncontextual hidden-variable model~\cite{kirby2020classical, kirby2021contextual, weaving2023stabilizer}. The reduced contextual subspace Hamiltonian is obtained as in Equation \eqref{eq:cs_proj}.

\begin{figure}[t]
    \centering
    \includegraphics[width=\linewidth]{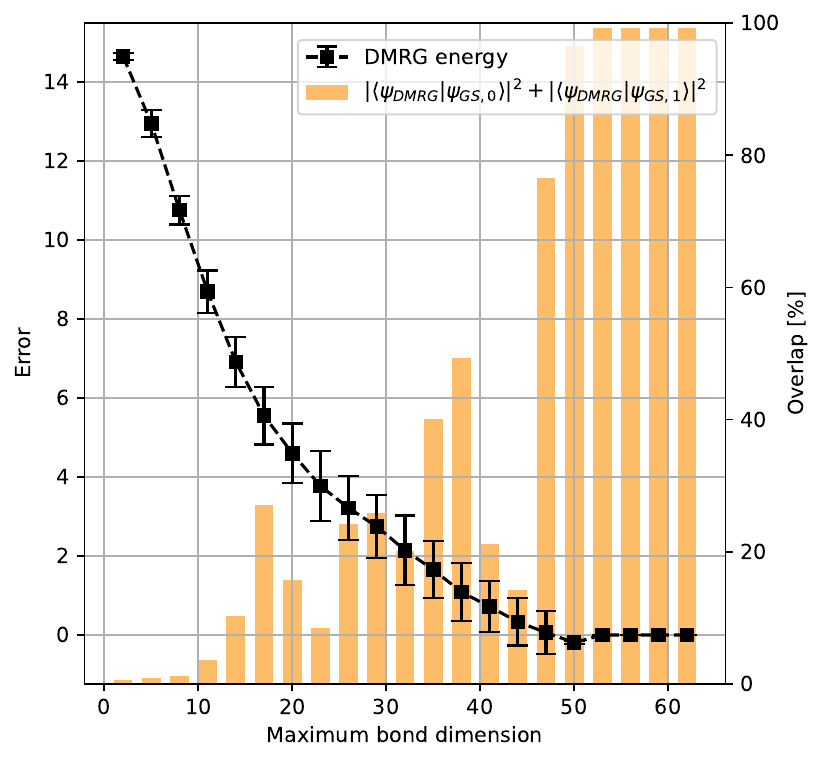}
    \caption{DMRG energy and ground state overlap for increasing bond dimension, with the average energy and standard deviation obtained over $300$ runs of DMRG with random shuffling of the Hamiltonian term order in the MPO construction. We plot the total overlap obtained as a sum of each eigenstate in the doubly degenerate ground space. }
    \label{fig:DMRG_energy}
\end{figure}

The success of this approach now lies with the choice of approximate wavefunction $\ket{\psi}$. For the present Kagome lattice Heisenberg model application, we utilize states obtained from a low bond dimension tensor network approach. For the Hamiltonian $H_{\mathrm{taper}}$ in Equation \eqref{H_taper}, we prepare a Matrix Product Operator (MPO) by summing the individual terms of $H_{\mathrm{taper}}$ represented as an MPO. During this summation, we truncate to a maximum bond dimension $D_{\mathrm{max}}$ to maintain scalability. Once prepared, we execute a Density-Matrix Renormalization Group (DMRG) calculation \mbox{(using the \texttt{quimb} Python package \cite{Gray2018})} to the MPO to prepare a wavefunction approximation $\ket{\psi_{\mathrm{DMRG}}^{D_{\mathrm{max}}}}$. There is some stochasticity to these calculations due to the use of random seeds. In Figure \ref{fig:DMRG_energy} we plot error with respect to the ground state energy \mbox{$E_0 = -18$} for DMRG calculations for increasing $D_{\mathrm{max}}$. On another axis we plot the sum of overlaps $|\braket{\psi_{\mathrm{DMRG}}^{D_{\mathrm{max}}} | \psi_{\mathrm{GS},0}}|^2 + |\braket{\psi_{\mathrm{DMRG}}^{D_{\mathrm{max}}} | \psi_{\mathrm{GS},1}}|^2$ against the two true wavefunctions in the doubly degenerate ground space.

\begin{figure*}[t]
    \centering
    \begin{subfigure}{0.45\linewidth}
    \includegraphics[width=\linewidth]{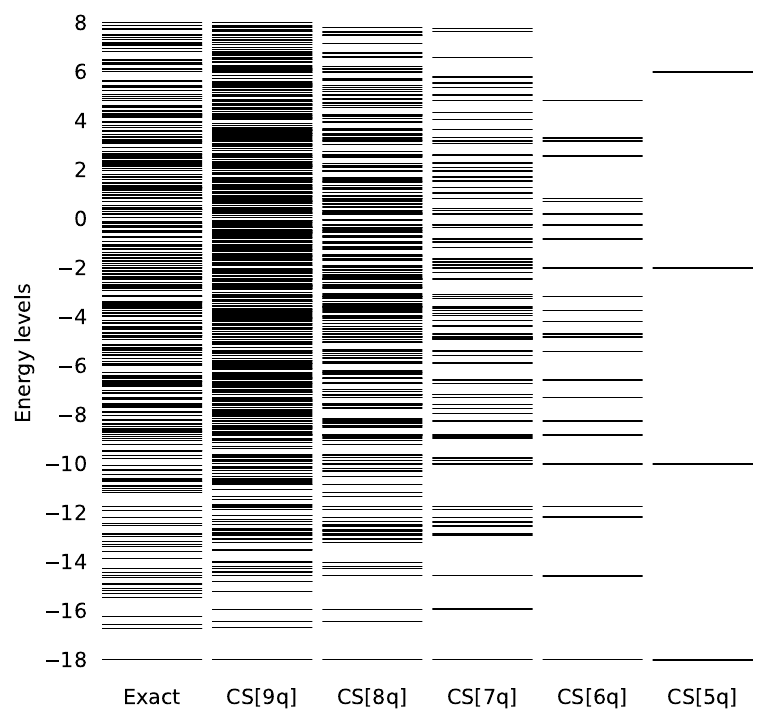}
    \caption{}
    \label{fig:spectra}
    \end{subfigure}
    \begin{subfigure}{0.45\linewidth}
    \includegraphics[width=\linewidth]{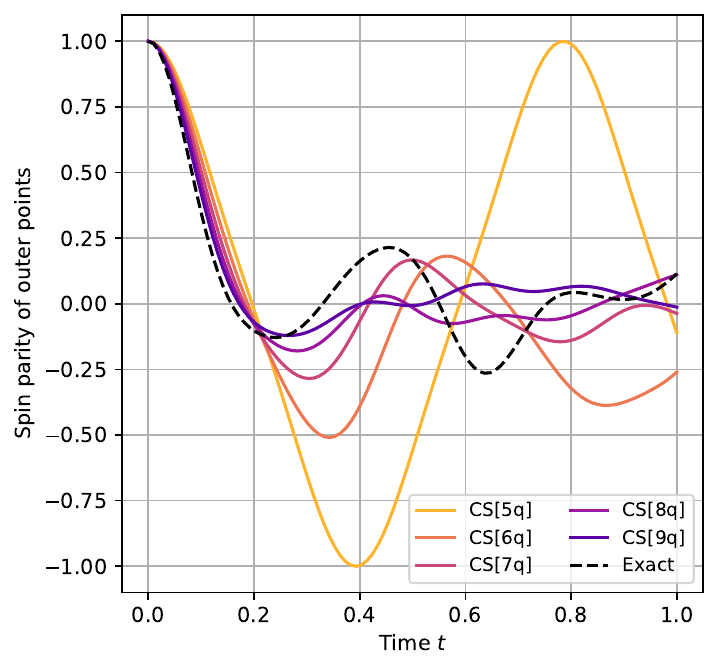}
    \caption{}
    \label{fig:time_evo}
    \end{subfigure}
    
    \caption{\textbf{(a)} Energy spectra computed via exact diagonalization of the $J=-1, h=0$ Heisenberg model over the Kagome lattice for different sized contextual subspaces, compared  alongside the full spectrum. The ground state energy $E_0=-18$ is preserved exactly in each subspace, since the stabilizer selection is biased towards the ground space using the methodology of Sections \ref{sec:approx} and \ref{sec:cs_bias}. However, note how higher energy levels, such as the first excited state, either degrade or vanish as the Hamiltonian is compressed; this can be counteracted by biasing the stabilizers towards the desired excited state. \textbf{(b)} We investigate how a compression of the energy spectrum can affect Hamiltonian time evolution, where we observe the evolution of spin parity, corresponding with the observable $\sigma_z^{6}\sigma_z^{7}\sigma_z^{8}\sigma_z^{9}\sigma_z^{10}\sigma_z^{11}$, for the outer points in the Kagome star lattice highlighted in Figure \ref{fig:kagome_lattice} and we see that smaller subspaces lose frequency components that are present in the full space.}
    
\end{figure*}

Respectable overlap is obtained for relatively low maximum bond dimensions; for example, with a limited $D_{\mathrm{max}}= 17$ the average overlap across the doubly generate ground space is $27.1\%$. Taking ${\ket\psi} = \ket{\psi_{\mathrm{DMRG}}^{17}}$ as the biasing wavefunction in the stabilizer selection heuristic described at the beginning of this section and culminating in Equation \eqref{eq:stab_select}, we then assess the quality of the resulting contextual subspaces. In Figure \ref{fig:spectra} we present the energy spectra of the antiferromagnetic Heisenberg model with $J=-1, h=0$ on a single cell of the Kagome lattice for contextual subspaces of varying size. Of note here is that the ground state energy is preserved exactly in each instance, whereas excited state energies either degrade or vanish as the subspace is reduced and the spectrum is correspondingly compressed. If we instead chose the wavefunction approximation $\ket{\psi}$ to bias towards a desired excited state, we could mitigate against that to bias the energy levels we wish to preserve under compression. However, for this demonstration we are interested specifically in ground state preparation. In Figure \ref{fig:time_evo} it is illustrated how this compression of the spectrum affects Hamiltonian time evolution.

\subsubsection{Effective Noncontextual Hamiltonian for the Kagome Lattice Heisenberg Model}\label{sec:noncon_ham}

With the goal of preparing the Kagome lattice ground state on a quantum computer, we choose to simulate a $5$-qubit contextual subspace Hamiltonian whose spectrum may be viewed in Figure \ref{fig:spectra}. This choice was made so that we may tile three concurrent circuit instances across the 16 qubits of the \texttt{ibmq\_guadalupe} superconducting device, effectively tripling the number of samples extracted from the quantum hardware in addition to averaging over noise; this is addressed in more detail later in Section \ref{sec:qsim}. 

Having applied the stabilizer heuristic of Section \ref{sec:cs_bias} to the tapered Hamiltonian $H_{\mathrm{taper}}$ in Equation \eqref{H_taper}, the explicit 5-qubit contextual subspace Hamiltonian is:
\begin{equation}\label{H_cs}
\begin{aligned}
    &H_{\mathrm{CS}} ={}
    - I + 7 \cdot \sigma_z^{(0)}\\ 
    & + (I + \sigma_z^{(0)}) (\sigma_z^{(1)} + \sigma_z^{(1)}\sigma_z^{(2)} + \sigma_z^{(2)}\sigma_z^{(3)} + \sigma_z^{(3)}\sigma_z^{(4)}) \\ 
    & - (I - \sigma_z^{(0)}) (\sigma_x^{(1)} + \sigma_x^{(2)} + \sigma_x^{(3)} - \sigma_x^{(4)} - \sigma_x^{(1)}\sigma_x^{(2)}\sigma_x^{(3)}\sigma_x^{(4)}).
\end{aligned}
\end{equation}
We also note the presence of a $\mathbb{Z}_2$-type symmetry $\sigma_z^{(0)}$. It is therefore possible to reduce the Hamiltonian to 4 qubits via Qubit Tapering, although we elect to retain this symmetry as it will be useful for later quantum error mitigation through Symmetry Verification (SV). Expressing the Hamiltonian of Equation \eqref{H_cs} in this way reveals projectors onto the $\pm1$-eigenspaces of this symmetry; indeed, we note it is the $-1$ eigenvalue that will permit the lowest energy due to the dominant $7 \cdot \sigma_z^{(0)}$ term, thus we could drop the $+1$-eigenspace projector and any corresponding terms from the Hamiltonian. However, we instead opt to extract a noncontextual component from this Hamiltonian, achieved by retaining any single term corresponding with the $+1$-projector, say $(I + \sigma_z^{(0)})\sigma_z^{(1)}\sigma_z^{(2)}$, while discarding the others. It is now possible to express the contextual subspace Hamiltonian as
\begin{equation}
\begin{aligned}
    &H_{\mathrm{CS}}^\prime ={}\\ &-I + 7 S_0 - S_2 + S_3 + S_0S_2 - S_0S_3 + S_1S_2S_3 - S_0S_1S_2S_3  \\
    & + C_0(-I + S_0 - S_1 + S_0S_1) \\
    & + C_1(I + S_0)
\end{aligned}
\end{equation}
where
\begin{equation}
\begin{aligned} 
    &S_0 = \sigma_z^{(0)}, S_1 = \sigma_x^{(1)} \sigma_x^{(2)}, S_2 = \sigma_x^{(3)}, S_3 = \sigma_x^{(4)}, \\
    &C_0 = \sigma_x^{(2)}, C_1 = \sigma_z^{(1)} \sigma_z^{(2)}.
\end{aligned}
\end{equation}
In particular, the Hamiltonian has been decomposed into a symmetry component plus a sum over commuting cliques such that terms across cliques pairwise anticommute. Calling $C_0, C_1$ the clique \textit{representatives}, since $\{C_0, C_1\}=0$ and the $S_i$ commute globally amongst terms of the Hamiltonian, $H_{\mathrm{CS}}^\prime$ constitutes a noncontextual Hamiltonian in line with the strict definition of Kirby \& Love \cite{kirby2019contextuality, kirby2020classical}. 

The energy spectrum is therefore parametrized by eigenvalue assignments to the symmetry elements $S_0, S_1, S_2, S_3$, specifying a symmetry \textit{sector}, and clique operator $C(r_0, r_1) = r_0C_0 + r_1C_1$ with $|r_0|^2+|r_1|^2=1$. We may search the symmetry sectors with $R_y$ rotations through binary relaxation of the eigenvalue assignments, since $R_y(\theta_i) \ket{0} = \cos(\theta_i/2)\ket{0} + \sin(\theta_i/2)\ket{1}$. Treatment of the clique operator is slightly different -- in solving the noncontextual problem we optimize over clique weights $r_0, r_1$ which in turn yields a rotation of the form in Unitary Partitioning \cite{izmaylov2019unitary, ralli2023unitary}, the specific form of which is found by exponentiating products of clique representatives $e^{\frac{\theta}{2} C_0C_1} = R_{zy}(\theta)$ with $C_0C_1 = -i \sigma_z^{(1)} \sigma_y^{(2)}$. Therefore, we are able to construct a simplistic ansatz
\begin{equation}\label{ansatz}
    \ket{\psi(\bm{\theta})} = e^{-i\frac{\theta_5}{2} \sigma_z^{(1)} \sigma_y^{(2)}} 
    \prod_{n=0}^{4} e^{-i\frac{\theta_n}{2} \sigma_y^{(n)}}
    \ket{\bm{0}},
\end{equation}
expressed as a circuit in Figure \ref{fig:NC_ansatz}, that is fully-expressible, given that it searches the noncontextual energy spectrum that we have deemed to capture the ground state.

\section{Methods}\label{sec:qsim}

Following the results of Section \ref{sec:noncon_ham}, we have a reduced 5-qubit contextual subspace Hamiltonian, given in Equation \eqref{H_cs}, whose ground state energy coincides exactly with that of the Heisenberg $XXX$ Hamiltonian with $J=-1, h=0$ over the 12-site single cell Kagome lattice, seen in Figure \ref{fig:kagome_lattice}. The Contextual Subspace method is available in the \texttt{symmer} package \cite{symmer2022}.  We moreover have and a simple ansatz circuit, provided in Equation \eqref{ansatz} and which is fully expressible. We are able to tile three instances of this ansatz circuit, shown in Figure \ref{fig:NC_ansatz}, across the 16-qubits available on \texttt{ibmq\_guadalupe}, tripling the number of circuit shots obtained and averaging over hardware noise, as depicted in Figures \ref{fig:parallel} and \ref{fig:guadalupe_topology}. Circuit construction, tiling and execution was implemented in \texttt{qiskit} \cite{Qiskit}.

The measurement overhead for our quantum experiments can be further reduced by implementing a Pauli measurement grouping strategy. Two Pauli strings $\bm{\sigma}_1, \bm{\sigma}_2$ are said to be qubit-wise commuting (QWC) if $[\sigma_1^{(n)}, \sigma_2^{(n)}] = 0 \;\; \forall n \in \mathbb{Z}_N$. This is useful as the measurement outcome for a collection of QWC Pauli strings may be inferred from only a single physical measurement \cite{verteletskyi2020measurement}. Conveniently, our contextual subspace Hamiltonian in Equation \eqref{H_cs} is expressible as a sum of just two QWC groups, thus reducing the measurement cost significantly. More sophisticated approaches can also be used, for example based on commuting \cite{yen2020measuring} or anticommuting \cite{izmaylov2019unitary, zhao2020measurement, ralli2023unitary} clique covers, but these require additional coherent resource.

Moving on to address the Variational Quantum Eigensolver (VQE) routine itself, we used both the Broyden–Fletcher–Goldfarb–Shanno (BFGS) \cite{fletcher2013practical} and Conjugate Gradient (CG) \cite{hestenes1952methods} algorithms implemented in \texttt{scipy} \cite{2020SciPy-NMeth} to perform the classical optimization over our six ansatz parameters given explicitly in Equation \eqref{ansatz} with objective function $E(\bm{\theta}) = \bra{\psi(\bm{\theta})} H_{\mathrm{CS}}\ket{\psi(\bm{\theta})}$.
The parameter gradients are computed in-hardware using the parameter shift rule \cite{parrish2019hybrid}:
\begin{equation}
    \frac{\partial}{\partial \theta_k} E(\bm{\theta}) = E\Big(\bm{\theta} + \frac{\pi}{4} \bm{e}_k\Big) - E\Big(\bm{\theta} - \frac{\pi}{4} \bm{e}_k \Big)
\end{equation}
where $\bm{e}_k$ is zero except in element $k$ where it is one.

In order to obtain the most accurate energy estimates possible, we deploy a Quantum Error Mitigation (QEM) strategy comprised of several techniques that we previously benchmarked for ground state preparation of the \ce{HCl} molecule \cite{weaving2023benchmarking}. The QEM methods utilized for this Kagome simulation are discussed in the proceeding subsections.

\subsubsection{Readout Error Mitigation}

Readout Error Mitigation (REM) focuses on rectifying errors that occur due to the measurement process in quantum experiments, treating earlier circuit operations as a black box. To address scalability challenges in constructing full assignment/transition matrices for large qubit systems, a ``tensored'' approach that estimates readout errors independently for each qubit can be adopted \cite{nation2021scalable, bravyi2021mitigating}. Specifically, tensored REM constructs single-qubit assignment matrices $A^{(n)}$, where $A^{(n)}_{i,j}$ is the observed probability of qubit $q_n$ being in state $\ket{i}$~and measuring $\ket{j}$. 
Then, the overall $N$-qubit assignment matrix is approximated by taking products of the single-qubit assignment matrices. This method can very successful under the assumption that readout errors are predominantly uncorrelated; this assumption was empirically validated on the \texttt{ibm\_kolkata} device, where the tensored approach closely matched results from full measurement calibration until intentional correlations were introduced by modifying readout pulse amplitudes \cite{nation2021scalable}. The \texttt{mthree} package provides tools for performing REM. An alternative approach that could be used and does not make assumptions on the absence of correlated measurements is Twirled Readout Error Extinction (TREX), leveraging the idea of twirled measurements \cite{van2022model}.

\subsubsection{Symmetry Verification}

In Symmetry Verification (SV) we discard quantum measurements that violate some known symmetry of the system \cite{bonet2018low, mcardle2019error, cai2021quantum}. Commuting operators share a common eigenbasis, so for a symmetry $S$ of a Hamiltonian $H$, meaning $[H,S]=0$, if $H\ket{\psi}=E\ket{\psi}$ then we must also have $S\ket{\psi}=s\ket{\psi}$ for some eigenvalue $s$ of $S$. In the case of $\mathbb{Z}_2$-type symmetry, we will have $s = \pm1$ and we may project onto valid measurements via
$\mathbb{P}=\frac{1}{2} (1\pm S)$. Therefore, we may throw away any measured bitstrings $\bm{m}$ such that $\pm S\ket{\bm{m}} \neq \ket{\bm{m}}$, i.e. that are not stabilizer by $\pm S$. This postselection procedure can often be implemented for non-$\mathbb{Z}_2$ symmetry as well, for example if a system obeys particle number symmetry, we can discard measurements with the wrong Hamming weight (the number of ones in a binary string). The reduced Hamiltonian in Equation \eqref{H_cs} possesses a single $\mathbb{Z}_2$-type symmetry $\sigma_z^{(0)}$ whose optimal eigenvalue assignment is $-1$. Therefore, we need only keep measurement outcomes $\ket{\bm{m}}$ that lie within the correct symmetry sector, by postselecting on those that are stabilized by $-\sigma_z^{(0)}$ (i.e. when qubit $q_0$ measures $m_0=1$).

\begin{figure}[t]
    \centering
    \begin{subfigure}{\linewidth}
    \resizebox{\linewidth}{!}{
    \input{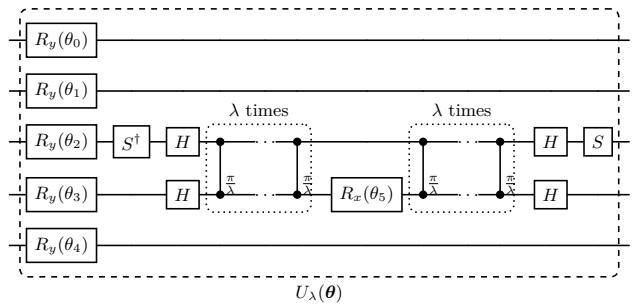}
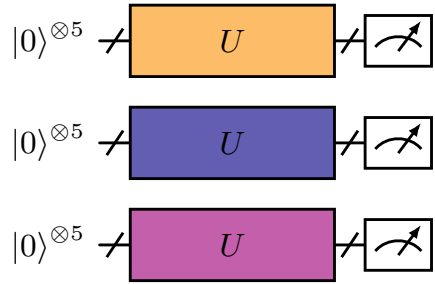
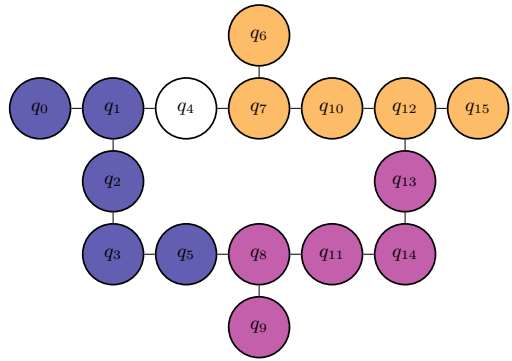
    }
    \caption{A 5-qubit ansatz circuit with noise amplification (controlled via the noise scaling factor $\lambda$) defined over the contextual subspace that searches the noncontextual energy landscape.}
    \label{fig:NC_ansatz}
    \end{subfigure}
    \begin{subfigure}{\linewidth}
    \resizebox{0.7\linewidth}{!}{
    \begin{quantikz}[row sep=3mm, column sep=3mm]
        \lstick{$\ket{0}^{\otimes 5}$} & \gate[style={fill=N2_plasma4!75}][2cm][7mm]{U} \qwbundle[alternate]{} & \meter{} \qwbundle[alternate]{} \\
        \lstick{$\ket{0}^{\otimes 5}$} & \gate[style={fill=N2_plasma0!65}][2cm][7mm]{U} \qwbundle[alternate]{} & \meter{} \qwbundle[alternate]{} \\
        \lstick{$\ket{0}^{\otimes 5}$} & \gate[style={fill=N2_plasma2!75}][2cm][7mm]{U} \qwbundle[alternate]{} & \meter{} \qwbundle[alternate]{} \\
    \end{quantikz}
    }
    \caption{Circuit tiling for improved sampling and noise averaging.}
    \label{fig:parallel}
    \end{subfigure}
    \begin{subfigure}{\linewidth}
    \resizebox{0.8\linewidth}{!}{
    \begin{tikzpicture}[auto, node distance=12mm,
 every node/.style={circle,thick,draw,minimum size=10mm}
 ]
\node (A) [            fill=N2_plasma0!65] {$q_0$};    
\node (B) [right of=A, fill=N2_plasma0!65] {$q_1$};    
\node (C) [right of=B] {$q_4$};
\node (D) [right of=C, fill=N2_plasma4!75] {$q_7$};    
\node (E) [right of=D, fill=N2_plasma4!75] {$q_{10}$}; 
\node (F) [right of=E, fill=N2_plasma4!75] {$q_{12}$};
\node (G) [right of=F, fill=N2_plasma4!75] {$q_{15}$};
\node (K) [below of=B, fill=N2_plasma0!65] {$q_2$};
\node (L) [below of=K, fill=N2_plasma0!65] {$q_3$};    
\node (M) [right of=L, fill=N2_plasma0!65] {$q_5$};    
\node (N) [right of=M, fill=N2_plasma2!75] {$q_8$};
\node (O) [right of=N, fill=N2_plasma2!75] {$q_{11}$}; 
\node (P) [right of=O, fill=N2_plasma2!75] {$q_{14}$}; 
\node (W) [below of=F, fill=N2_plasma2!75] {$q_{13}$}; 
\node (X) [below of=N, fill=N2_plasma2!75] {$q_9$};
\node (Z) [above of=D, fill=N2_plasma4!75] {$q_6$};


\draw (A) edge (B); \draw (B) edge (C); \draw (C) edge (D); \draw (D) edge (E); \draw (E) edge (F); 
\draw (F) edge (G); \draw (B) edge (K); \draw (K) edge (L); \draw (L) edge (M); \draw (M) edge (N); \draw (N) edge (O); \draw (O) edge (P); \draw (F) edge (W); \draw (D) edge (Z);
\draw (P) edge (W); \draw (N) edge (X);
\end{tikzpicture}
    }
    \caption{The \texttt{ibmq\_guadalupe} Falcon r4P chip ``heavy-hex'' topology.}
    \label{fig:guadalupe_topology}
    \end{subfigure}

    
    
    \caption{Routing the noncontextual ansatz circuit onto the available device topology.}
    \label{fig:top_map}
\end{figure}


\subsubsection{Zero Noise Extrapolation}

The idea underpinning Zero Noise Extrapolation (ZNE) is that, if we can controllably introduce noise into a quantum system, we can analyze its behaviour and make some inference on what the experimentally impenetrable ``zero-noise'' regime looks like. We can amplify noise arising from some selection of gates in a quantum circuit, for example discretely through a unitary folding scheme or continuously in the time domain via gate stretching. Either way, we obtain noise-scaled energy estimates that one may use for extrapolation to the hypothetical point of zero-noise \cite{li2017efficient, temme2017error, endo2018practical, kandala2019error, giurgica2020digital, he2020zero, mari2021extending}. In our implementation we choose to amplify CNOT noise by decomposing over a product of controlled phase gates: for control qubit $c$ and target $t$, \begin{equation}
    \mathrm{CNOT}_{c,t} = \mathrm{Had}^{(t)} \prod_{n=1}^{\lambda} \mathrm{CPhase}_{c,t}\Big(\frac{\pi}{\lambda}\Big) \mathrm{Had}^{(t)}    
\end{equation}
where $\lambda$ is a noise/gain scaling parameter and $\mathrm{Had}$ is the Hadamard gate. An example of this can be viewed in Figure \ref{fig:NC_ansatz}, where the parametrized circuit $U_{\lambda}(\bm{\theta})$ contains $2\lambda$ CPhase gates, noting upon transpilation each CPhase will be mapped onto a pair of CNOTs, resulting in $4\lambda$ CNOT gates.

After evaluating several noise-amplified expectation values \mbox{$E_{\lambda}(\bm{\theta}) = \bra{\bm{0}} U_{\lambda}(\bm{\theta})^\dagger H_{\mathrm{CS}} U_{\lambda}(\bm{\theta}) \ket{\bm{0}}$} for a set of noise factors (here we take $\lambda \in \{1,2,3,4\}$) on a quantum device, we perform an exponential fitting routine \mbox{$f(\lambda; \alpha,\beta,\gamma) = e^{\alpha\lambda}e^\beta+\gamma$}, so that $\log [f(\lambda)-\gamma]$ is linear. The fitting curve is optimized by minimizing a weighted sum of squared residuals
\begin{equation}\label{rss}
    \mathrm{RSS}(\alpha, \beta, \gamma) = \sum_{\lambda} \bigg[\frac{E_{\lambda}(\bm{\theta}) - f(\lambda; \alpha,\beta,\gamma)}{\sigma(\lambda)}\bigg]^2
\end{equation} 
where $\sigma(\lambda)$ is the experimentally-obtained standard deviation at noise amplification factor $\lambda$. The inclusion of a weighting factor $\frac{1}{\sigma(\lambda)^2}$ (inverse variance) biases the fitting towards data of low variance, while highly-varying points are penalized.

\section{Simulation Results}\label{sec:qsim_res}

\begin{figure}[b!]
    \centering
    \includegraphics[width=\linewidth]{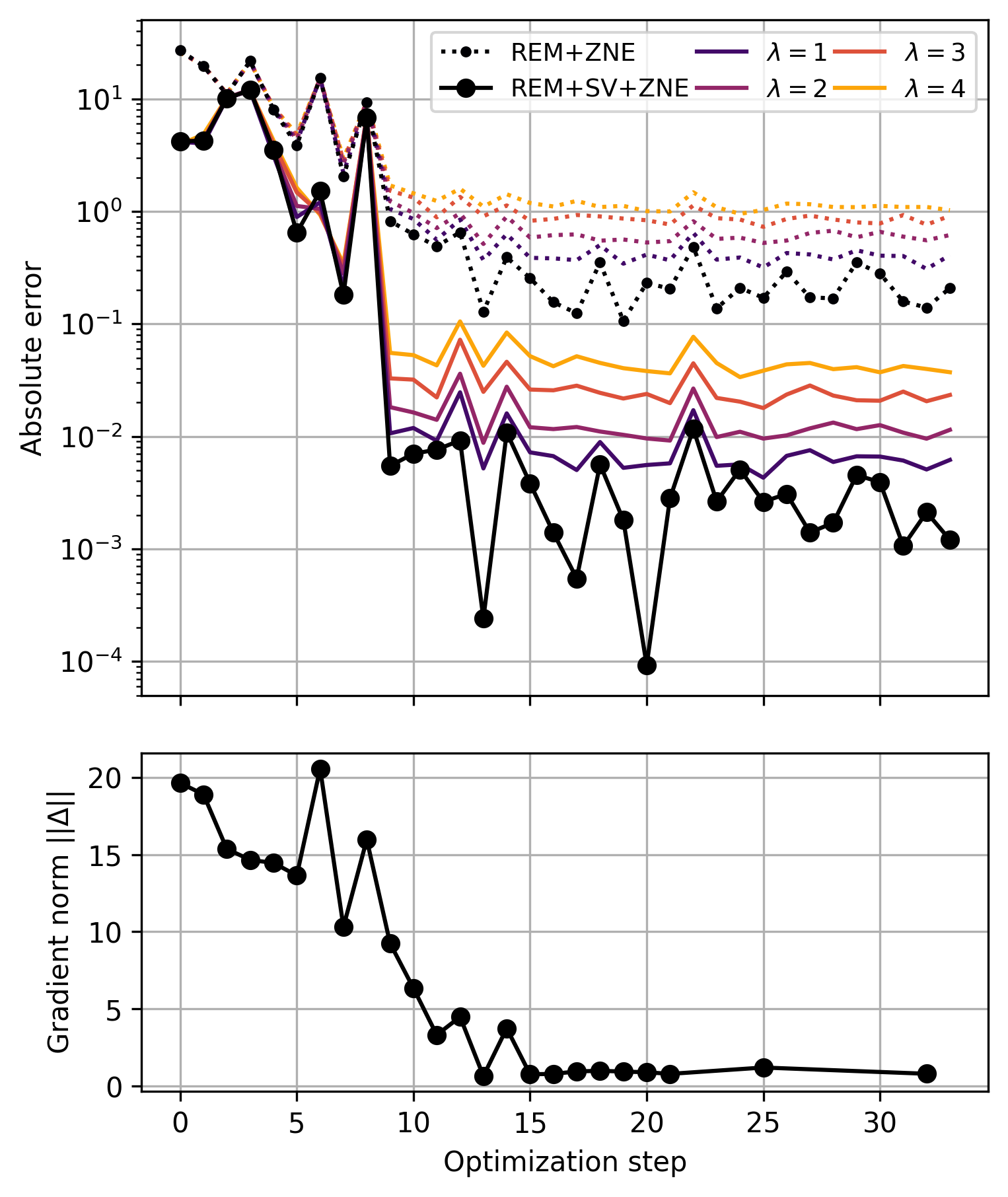}
    \caption{VQE optimization trace over the ansatz circuit in Figure \ref{fig:NC_ansatz} with respect to a 5-qubit contextual subspace Hamiltonian, given in Equation \eqref{H_cs}, for a single cell Kagome lattice system. Curves of different colour relate to noise amplification factors $\lambda \in \{1,2,3,4\}$, dotted lines indicate that just REM was applied, while solid lines used REM+SV in the QEM strategy. The black curves show the results from an exponential fit minimizing the sum of squared residuals in Equation \eqref{rss} for ZNE.}
    \label{fig:vqe_trace}
\end{figure}

We now apply all the methods introduced in this work to a Variational Quantum Eigensolver (VQE) simulation run on \texttt{ibm\_quadalupe} of the 5-qubit contextual subspace Hamiltonian in Equation \eqref{H_cs}, describing a single cell of the Kagome lattice depicted in Figure \ref{fig:kagome_lattice}, with the ansatz circuit in Figure \ref{fig:NC_ansatz} consisting of six rotation parameters. In Figure \ref{fig:vqe_trace} we present the optimization trace for a noise-amplified VQE routine; different colours indicate the level of noise amplification for factors $\lambda\in\{1,2,3,4\}$, while the black curves are the extrapolated estimates according to an exponential fit that minimizes the sum of squared residuals in Equation \eqref{rss}.  We see the addition of SV error mitigation on top of REM suppresses the energy errors by $1-2$ orders of magnitude. In the VQE routine itself the ZNE fit is unweighted, meaning $\sigma(\lambda)=1$; we could have estimated standard deviations through statistical bootstrapping or by running additional quantum experiments. 

The CG optimizer aims to minimize the energy and parameter gradient, the latter of which is also evaluated using ZNE on the quantum hardware. This adds a significant number of shots to the overall budget, which could be mitigated by only using ZNE on the energy estimates and not the individual partial derivatives with respect to circuit parameters. The simulation consists of $N_{\mathrm{energy}} = 34$ energy estimates and $N_{\mathrm{grad}} = 24$ gradient evaluations over $N_{\mathrm{param}} = 6$ parameters. Each estimate requires $N_{\mathrm{ZNE}} = 4$ noise amplified expectation value calculations over $N_{\mathrm{QWC}} = 2$ QWC groups. For the VQE routine presented in Figure \ref{fig:vqe_trace} we set $N_{\mathrm{shots}}=2^{13}=8192$, although we also ran successful simulations for $N_{\mathrm{shots}} \in \{256, 1024, 4096, 8192\}$ with both BFGS and conjugate gradient (CG) classical optimizers. Altogether, the total number of circuit executions performed on \texttt{ibmq\_guadalupe} is
\begin{equation}
\begin{aligned}
    N_{\mathrm{budget}} ={} & N_{\mathrm{shots}}  N_{\mathrm{QWC}}  N_{\mathrm{ZNE}}  (N_{\mathrm{energy}} + 2  N_{\mathrm{grad}}  N_{\mathrm{param}}) \\
    ={} & N_{\mathrm{shots}} \times 2576 \\
    \approx & 2.11 \times 10^7.
\end{aligned}
\end{equation}
If we had not applied ZNE to the parameter gradient evaluation, we would have used $N_{\mathrm{budget}} \approx 6.95 \times 10^6$ shots for the whole simulation.

\begin{figure}[b!]
    \centering
    \includegraphics[width=\linewidth]{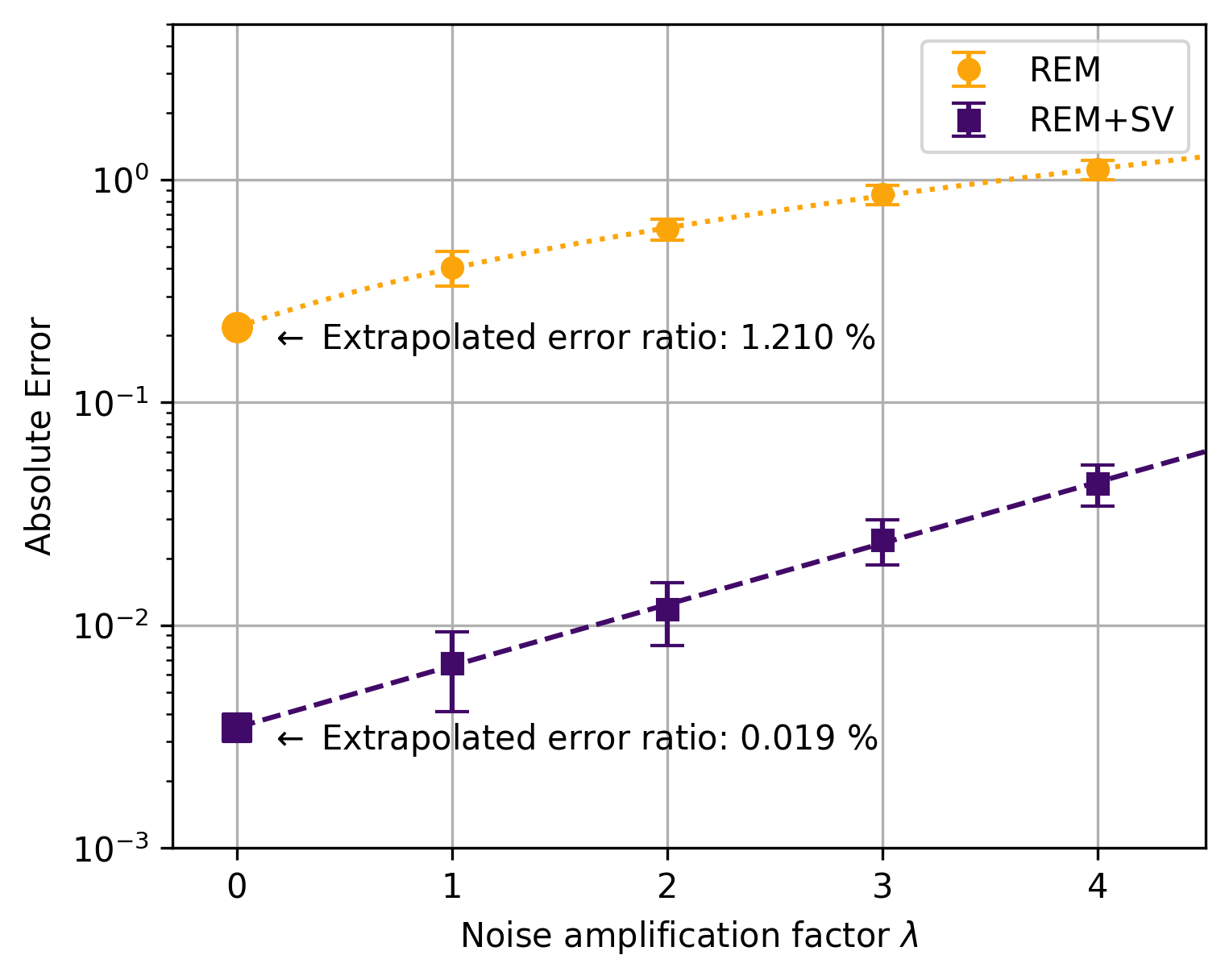}
    \caption{Final noise-amplified energy estimates and exponential weighted least squares fitting to obtain an accurate zero-noise extrapolated energy. Explicit values are provided in Table \ref{tab:zne_values}; standard deviations for REM+SV are lower than REM but the log scale obscures this.}
    \label{fig:zne_weighted}
\end{figure}

In Figure \ref{fig:vqe_trace} optimization converges at step $15$, after which point we compute several more steps at a fixed circuit parametrization to allow the evaluation of energy variances. This allows us to then perform \textit{weighted} ZNE to produce a final energy estimate for our Kagome problem. A weighted least squares (WLS) fitting scheme provides more reliable extrapolation since it penalizes highly varying data and biases towards those of low variance. This is achieved by evaluating the standard deviations $\sigma(\lambda)$ over the converged energy data for inclusion in the sum of squared residuals of Equation \ref{rss}. Each residual becomes weighted by the inverse variance $\frac{1}{\sigma(\lambda)^2}$ which is large for low variance and vice versa, thus placing greater importance on the residuals corresponding with data of lower variance. In Figure \ref{fig:zne_weighted} we present the final ZNE energy estimate and error ratio using WLS, both with and without application of SV. The error ratio as a percentage is given by $\frac{E_{\mathrm{ZNE}}-E_0}{E_0} \times 100 \%$. Using a QEM strategy of REM+ZNE yields a final error ratio of $1.210\%$, while the addition of SV brings the ratio down to $0.019\%$ with REM+SV+ZNE. Furthermore, the standard deviation is suppressed by SV, consistent with our previous QEM benchmarking efforts \cite{weaving2023benchmarking}. The log scale in Figure \ref{fig:zne_weighted} makes it difficult to compare standard deviations, so the explicit values are provided in Table \ref{tab:zne_values}.
 
\begin{table}[t!]
    \centering
    \begin{tabularx}{\linewidth}{XXXXX} \toprule
     & \multicolumn{2}{c}{REM} & \multicolumn{2}{c}{REM+SV} \\
    \cmidrule(r{-.5pt}){2-3} \cmidrule(l{10pt}){4-5}
    $\lambda$ & $E(\lambda)$ & $\sigma(\lambda)$ & $E(\lambda)$ & $\sigma(\lambda)$  \\ \midrule
     $1$      & -17.59609    &  0.07113          & -17.99325    &  0.00266           \\
     $2$      & -17.40111    &  0.06654          & -17.98822    &  0.00368           \\
     $3$      & -17.13956    &  0.08612          & -17.97595    &  0.00551           \\
     $4$      & -16.88709    &  0.10957          & -17.95655    &  0.00924           \\

    \bottomrule
    \end{tabularx}
    \caption{The noise amplified data in Figure \ref{fig:zne_weighted}, comparing estimates obtained with and without Symmetry Verification (SV) applied on top of Readout Error Mitigation (REM).}
    \label{tab:zne_values}
\end{table}

\begin{figure}[t!]
    \centering
    \begin{subfigure}{.49\linewidth}
        \includegraphics[width=\linewidth]{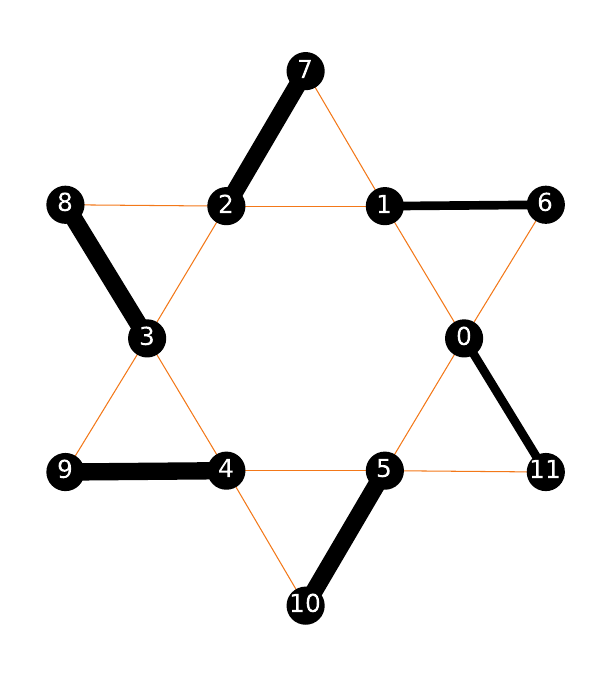}
        \caption{CS[5q]}
        \label{fig:pin_cs}
    \end{subfigure}
    \begin{subfigure}{.49\linewidth}
        \includegraphics[width=\linewidth]{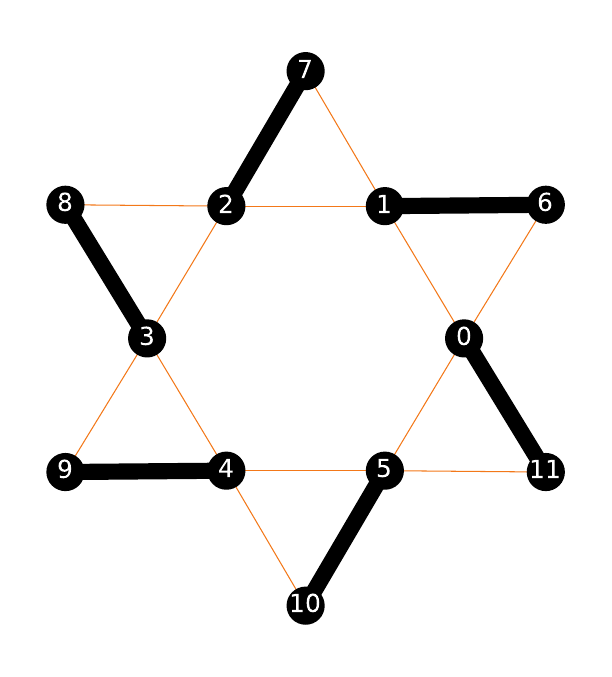}
        \caption{Exact}
        \label{fig:pin_exact}
    \end{subfigure}
    
    \caption{Ground state comparison of Kagome spin couplings between \textbf{(a)}~the wavefunction obtained from running a noisy VQE simulation in a 5-qubit contextual subspace and \textbf{(b)} the exact wavefunction. We observe the distinctive pinwheel motif that can arise as a singlet ground state from this frustrated lattice geometry \cite{matan2010pinwheel}, with the doubly degenerate ground state obtained by a reflection through the vertical axis. The black line thickness indicates the mutual information between spin sites, noting the couplings $(1,6)$ and $(0,11)$ are weaker in the VQE simulation of \textbf{(a)} compared with \textbf{(b)}}
    \label{fig:pinwheel}
\end{figure}

Finally, in Figure \ref{fig:pinwheel} we present the spin couplings present in the final Contextual Subspace VQE wavefunction as compared with the true wavefunction obtained via exact diagonalization. The thickness of the black lines indicates the pairwise mutual information between spin sites and we observe the distinctive pinwheel pattern \cite{matan2010pinwheel}. This arises here due to spin frustration in the antiferromagnetic regime and the geometrical structure of the Kagome lattice. The spins should form singlets $\frac{1}{\sqrt{2}}\big(\ket{01}-\ket{10}\big)$ as in the exact result of Figure \ref{fig:pin_exact}, while in Figure \ref{fig:pin_cs} they are approximate up to some small degree of coupling with other sites. This can be seen between spins $(1,6)$ and $(0,11)$ where the mutual information is lower, indicated by the lines being thinner.

\section{Conclusion}

In this work we calculated highly accurate energy estimates for the antiferromagnetic, field-free Heisenberg model on a single cell of the Kagome lattice on quantum computer. The Kagome lattice has a geometrical structure from which exotic physical properties emerge and is a promising platform for potentially realizing a quantum spin liquid material. These estimates were obtained through the application of many contemporary techniques in quantum computing. We deployed a qubit subspace approach that leveraged the Contextual Subspace framework, built on a foundation of strong measurement contextuality. The subspace selection was biased using a approximate symmetry generator algorithm that utilizes the symplectic representation of the Pauli group. This was combined with Density Matrix Renormalization Group calculations of a low bond dimension, while the resulting subspace energies outperform the DMRG energy at a specified bond dimension. We then deployed the Variational Quantum Eigensolver with an error mitigation strategy comprised of Readout Error Mitigation, Symmetry Verification and Zero Noise Extrapolation with exponential Weighted Least Squares fitting. This methodology resulted in very low error in the energy estimates produced from the quantum computer, with a final error ratio of just $0.019\%$.

In future work we would like to investigate whether the Contextual Subspace approach continues to be effective as it is scaled to larger lattices. For this single-cell simulation we identified a noncontextual effective Hamiltonian whose ground state coincides exactly with the full system. It is unlikely that this will be possible for larger lattices, but it could be that near-noncontextual models still provide benefit and we may further leverage the Contextual Subspace methodology to yield more accurate quantum corrections to a noncontextual effective model. Another problem that will need addressing as we scale up is how to optimally map the spin lattice onto the hardware graph, which is of particular concern for superconducting systems. This is further nuanced by the fact that in projecting onto a contextual subspace, the sequence of rotations that rotate stabilizers onto distinct qubit positions mixes spin sites together. This means that each subspace qubit can describe a linear combination of spins, potentially resulting in Pauli terms of greater nonlocality than the original spin model. Finally, another route to pursue is the place of Contextual Subspace methods in the fault tolerant regime of quantum computing, since it is fully compatible with the longer term vision of Quantum Phase Estimation (QPE) and Quantum Signal Processing (QSP) techniques.

\section*{Acknowledgements}
T.W. and V.W. acknowledge funding from \mbox{EPSRC}, grant numbers EP/S021582/1 and EP/S516090/1. T.W. is additionally supported by \mbox{CBKSciCon} Ltd.. A.R. and P.J.L. are supported by the NSF STAQ project, award number PHY-1818914. P.V.C. is grateful for funding from the European Commission for VECMA, grant number 800925, and EPSRC for SEAVEA, grant number EP/W007711/1. Access to hardware was facilitated through the IBM Quantum Open Science Prize 2023, for which we are thankful to have been awarded first place.

\newpage

\section*{Author Contributions}
T.W., A.R. and V.W. formed a team for the IBM Quantum Open Science Prize 2023. T.W. developed the quantum algorithms and error mitigation strategy, in addition to designing and deploying the quantum experiments on IBM hardware and conducted the subsequent data analysis and post-processing. T.W. and A.R. developed the Contextual Subspace stabilizer methodology adopted here and the \texttt{symmer} software. V.W. wrote code for the tensor network techniques used in this work.

\vspace{5mm}

\bibliography{main}

\begin{thebibliography}{54}%
\makeatletter
\providecommand \@ifxundefined [1]{%
 \@ifx{#1\undefined}
}%
\providecommand \@ifnum [1]{%
 \ifnum #1\expandafter \@firstoftwo
 \else \expandafter \@secondoftwo
 \fi
}%
\providecommand \@ifx [1]{%
 \ifx #1\expandafter \@firstoftwo
 \else \expandafter \@secondoftwo
 \fi
}%
\providecommand \natexlab [1]{#1}%
\providecommand \enquote  [1]{``#1''}%
\providecommand \bibnamefont  [1]{#1}%
\providecommand \bibfnamefont [1]{#1}%
\providecommand \citenamefont [1]{#1}%
\providecommand \href@noop [0]{\@secondoftwo}%
\providecommand \href [0]{\begingroup \@sanitize@url \@href}%
\providecommand \@href[1]{\@@startlink{#1}\@@href}%
\providecommand \@@href[1]{\endgroup#1\@@endlink}%
\providecommand \@sanitize@url [0]{\catcode `\\12\catcode `\$12\catcode `\&12\catcode `\#12\catcode `\^12\catcode `\_12\catcode `\%12\relax}%
\providecommand \@@startlink[1]{}%
\providecommand \@@endlink[0]{}%
\providecommand \url  [0]{\begingroup\@sanitize@url \@url }%
\providecommand \@url [1]{\endgroup\@href {#1}{\urlprefix }}%
\providecommand \urlprefix  [0]{URL }%
\providecommand \Eprint [0]{\href }%
\providecommand \doibase [0]{https://doi.org/}%
\providecommand \selectlanguage [0]{\@gobble}%
\providecommand \bibinfo  [0]{\@secondoftwo}%
\providecommand \bibfield  [0]{\@secondoftwo}%
\providecommand \translation [1]{[#1]}%
\providecommand \BibitemOpen [0]{}%
\providecommand \bibitemStop [0]{}%
\providecommand \bibitemNoStop [0]{.\EOS\space}%
\providecommand \EOS [0]{\spacefactor3000\relax}%
\providecommand \BibitemShut  [1]{\csname bibitem#1\endcsname}%
\let\auto@bib@innerbib\@empty
\bibitem [{\citenamefont {Balents}(2010)}]{balents2010spin}%
  \BibitemOpen
  \bibfield  {author} {\bibinfo {author} {\bibfnamefont {L.}~\bibnamefont {Balents}},\ }\bibfield  {title} {\bibinfo {title} {Spin liquids in frustrated magnets},\ }\href {https://doi.org/https://doi.org/10.1038/nature08917} {\bibfield  {journal} {\bibinfo  {journal} {nature}\ }\textbf {\bibinfo {volume} {464}},\ \bibinfo {pages} {199} (\bibinfo {year} {2010})}\BibitemShut {NoStop}%
\bibitem [{\citenamefont {Nikoli{\'c}}\ and\ \citenamefont {Senthil}(2005)}]{nikolic2005theory}%
  \BibitemOpen
  \bibfield  {author} {\bibinfo {author} {\bibfnamefont {P.}~\bibnamefont {Nikoli{\'c}}}\ and\ \bibinfo {author} {\bibfnamefont {T.}~\bibnamefont {Senthil}},\ }\bibfield  {title} {\bibinfo {title} {Theory of the kagome lattice ising antiferromagnet in weak transverse fields},\ }\href {https://doi.org/10.1103/PhysRevB.71.024401} {\bibfield  {journal} {\bibinfo  {journal} {Physical Review B—Condensed Matter and Materials Physics}\ }\textbf {\bibinfo {volume} {71}},\ \bibinfo {pages} {024401} (\bibinfo {year} {2005})}\BibitemShut {NoStop}%
\bibitem [{\citenamefont {Matan}\ \emph {et~al.}(2010)\citenamefont {Matan}, \citenamefont {Ono}, \citenamefont {Fukumoto}, \citenamefont {Sato}, \citenamefont {Yamaura}, \citenamefont {Yano}, \citenamefont {Morita},\ and\ \citenamefont {Tanaka}}]{matan2010pinwheel}%
  \BibitemOpen
  \bibfield  {author} {\bibinfo {author} {\bibfnamefont {K.}~\bibnamefont {Matan}}, \bibinfo {author} {\bibfnamefont {T.}~\bibnamefont {Ono}}, \bibinfo {author} {\bibfnamefont {Y.}~\bibnamefont {Fukumoto}}, \bibinfo {author} {\bibfnamefont {T.~J.}\ \bibnamefont {Sato}}, \bibinfo {author} {\bibfnamefont {J.}~\bibnamefont {Yamaura}}, \bibinfo {author} {\bibfnamefont {M.}~\bibnamefont {Yano}}, \bibinfo {author} {\bibfnamefont {K.}~\bibnamefont {Morita}},\ and\ \bibinfo {author} {\bibfnamefont {H.}~\bibnamefont {Tanaka}},\ }\bibfield  {title} {\bibinfo {title} {Pinwheel valence-bond solid and triplet excitations in the two-dimensional deformed kagome lattice},\ }\href {https://doi.org/10.1038/nphys1761} {\bibfield  {journal} {\bibinfo  {journal} {Nature Physics}\ }\textbf {\bibinfo {volume} {6}},\ \bibinfo {pages} {865} (\bibinfo {year} {2010})}\BibitemShut {NoStop}%
\bibitem [{\citenamefont {Han}\ \emph {et~al.}(2012)\citenamefont {Han}, \citenamefont {Helton}, \citenamefont {Chu}, \citenamefont {Nocera}, \citenamefont {Rodriguez-Rivera}, \citenamefont {Broholm},\ and\ \citenamefont {Lee}}]{han2012fractionalized}%
  \BibitemOpen
  \bibfield  {author} {\bibinfo {author} {\bibfnamefont {T.-H.}\ \bibnamefont {Han}}, \bibinfo {author} {\bibfnamefont {J.~S.}\ \bibnamefont {Helton}}, \bibinfo {author} {\bibfnamefont {S.}~\bibnamefont {Chu}}, \bibinfo {author} {\bibfnamefont {D.~G.}\ \bibnamefont {Nocera}}, \bibinfo {author} {\bibfnamefont {J.~A.}\ \bibnamefont {Rodriguez-Rivera}}, \bibinfo {author} {\bibfnamefont {C.}~\bibnamefont {Broholm}},\ and\ \bibinfo {author} {\bibfnamefont {Y.~S.}\ \bibnamefont {Lee}},\ }\bibfield  {title} {\bibinfo {title} {Fractionalized excitations in the spin-liquid state of a kagome-lattice antiferromagnet},\ }\href {https://doi.org/https://doi.org/10.1038/nature11659} {\bibfield  {journal} {\bibinfo  {journal} {Nature}\ }\textbf {\bibinfo {volume} {492}},\ \bibinfo {pages} {406} (\bibinfo {year} {2012})}\BibitemShut {NoStop}%
\bibitem [{\citenamefont {Peruzzo}\ \emph {et~al.}(2014)\citenamefont {Peruzzo}, \citenamefont {McClean}, \citenamefont {Shadbolt}, \citenamefont {Yung}, \citenamefont {Zhou}, \citenamefont {Love}, \citenamefont {Aspuru-Guzik},\ and\ \citenamefont {O’Brien}}]{peruzzo2014variational}%
  \BibitemOpen
  \bibfield  {author} {\bibinfo {author} {\bibfnamefont {A.}~\bibnamefont {Peruzzo}}, \bibinfo {author} {\bibfnamefont {J.}~\bibnamefont {McClean}}, \bibinfo {author} {\bibfnamefont {P.}~\bibnamefont {Shadbolt}}, \bibinfo {author} {\bibfnamefont {M.-H.}\ \bibnamefont {Yung}}, \bibinfo {author} {\bibfnamefont {X.-Q.}\ \bibnamefont {Zhou}}, \bibinfo {author} {\bibfnamefont {P.~J.}\ \bibnamefont {Love}}, \bibinfo {author} {\bibfnamefont {A.}~\bibnamefont {Aspuru-Guzik}},\ and\ \bibinfo {author} {\bibfnamefont {J.~L.}\ \bibnamefont {O’Brien}},\ }\bibfield  {title} {\bibinfo {title} {A variational eigenvalue solver on a photonic quantum processor},\ }\href {https://doi.org/10.1038/ncomms5213} {\bibfield  {journal} {\bibinfo  {journal} {Nature communications}\ }\textbf {\bibinfo {volume} {5}},\ \bibinfo {pages} {1} (\bibinfo {year} {2014})}\BibitemShut {NoStop}%
\bibitem [{\citenamefont {Bosse}\ and\ \citenamefont {Montanaro}(2022)}]{bosse2022probing}%
  \BibitemOpen
  \bibfield  {author} {\bibinfo {author} {\bibfnamefont {J.~L.}\ \bibnamefont {Bosse}}\ and\ \bibinfo {author} {\bibfnamefont {A.}~\bibnamefont {Montanaro}},\ }\bibfield  {title} {\bibinfo {title} {Probing ground-state properties of the kagome antiferromagnetic heisenberg model using the variational quantum eigensolver},\ }\href {https://doi.org/10.1103/PhysRevB.105.094409} {\bibfield  {journal} {\bibinfo  {journal} {Physical Review B}\ }\textbf {\bibinfo {volume} {105}},\ \bibinfo {pages} {094409} (\bibinfo {year} {2022})}\BibitemShut {NoStop}%
\bibitem [{\citenamefont {Kattem{\"o}lle}\ and\ \citenamefont {Van~Wezel}(2022)}]{kattemolle2022variational}%
  \BibitemOpen
  \bibfield  {author} {\bibinfo {author} {\bibfnamefont {J.}~\bibnamefont {Kattem{\"o}lle}}\ and\ \bibinfo {author} {\bibfnamefont {J.}~\bibnamefont {Van~Wezel}},\ }\bibfield  {title} {\bibinfo {title} {Variational quantum eigensolver for the heisenberg antiferromagnet on the kagome lattice},\ }\href@noop {} {\bibfield  {journal} {\bibinfo  {journal} {Physical Review B}\ }\textbf {\bibinfo {volume} {106}},\ \bibinfo {pages} {214429} (\bibinfo {year} {2022})}\BibitemShut {NoStop}%
\bibitem [{\citenamefont {Kirby}\ \emph {et~al.}(2021)\citenamefont {Kirby}, \citenamefont {Tranter},\ and\ \citenamefont {Love}}]{kirby2021contextual}%
  \BibitemOpen
  \bibfield  {author} {\bibinfo {author} {\bibfnamefont {W.~M.}\ \bibnamefont {Kirby}}, \bibinfo {author} {\bibfnamefont {A.}~\bibnamefont {Tranter}},\ and\ \bibinfo {author} {\bibfnamefont {P.~J.}\ \bibnamefont {Love}},\ }\bibfield  {title} {\bibinfo {title} {Contextual subspace variational quantum eigensolver},\ }\href {https://doi.org/10.22331/q-2021-05-14-456} {\bibfield  {journal} {\bibinfo  {journal} {Quantum}\ }\textbf {\bibinfo {volume} {5}},\ \bibinfo {pages} {456} (\bibinfo {year} {2021})}\BibitemShut {NoStop}%
\bibitem [{\citenamefont {Weaving}\ \emph {et~al.}(2023{\natexlab{a}})\citenamefont {Weaving}, \citenamefont {Ralli}, \citenamefont {Kirby}, \citenamefont {Tranter}, \citenamefont {Love},\ and\ \citenamefont {Coveney}}]{weaving2023stabilizer}%
  \BibitemOpen
  \bibfield  {author} {\bibinfo {author} {\bibfnamefont {T.}~\bibnamefont {Weaving}}, \bibinfo {author} {\bibfnamefont {A.}~\bibnamefont {Ralli}}, \bibinfo {author} {\bibfnamefont {W.~M.}\ \bibnamefont {Kirby}}, \bibinfo {author} {\bibfnamefont {A.}~\bibnamefont {Tranter}}, \bibinfo {author} {\bibfnamefont {P.~J.}\ \bibnamefont {Love}},\ and\ \bibinfo {author} {\bibfnamefont {P.~V.}\ \bibnamefont {Coveney}},\ }\bibfield  {title} {\bibinfo {title} {A stabilizer framework for the contextual subspace variational quantum eigensolver and the noncontextual projection ansatz},\ }\href {https://doi.org/10.1021/acs.jctc.2c00910} {\bibfield  {journal} {\bibinfo  {journal} {Journal of Chemical Theory and Computation}\ }\textbf {\bibinfo {volume} {19}},\ \bibinfo {pages} {808} (\bibinfo {year} {2023}{\natexlab{a}})}\BibitemShut {NoStop}%
\bibitem [{\citenamefont {Ralli}\ \emph {et~al.}(2023)\citenamefont {Ralli}, \citenamefont {Weaving}, \citenamefont {Tranter}, \citenamefont {Kirby}, \citenamefont {Love},\ and\ \citenamefont {Coveney}}]{ralli2023unitary}%
  \BibitemOpen
  \bibfield  {author} {\bibinfo {author} {\bibfnamefont {A.}~\bibnamefont {Ralli}}, \bibinfo {author} {\bibfnamefont {T.}~\bibnamefont {Weaving}}, \bibinfo {author} {\bibfnamefont {A.}~\bibnamefont {Tranter}}, \bibinfo {author} {\bibfnamefont {W.~M.}\ \bibnamefont {Kirby}}, \bibinfo {author} {\bibfnamefont {P.~J.}\ \bibnamefont {Love}},\ and\ \bibinfo {author} {\bibfnamefont {P.~V.}\ \bibnamefont {Coveney}},\ }\bibfield  {title} {\bibinfo {title} {Unitary partitioning and the contextual subspace variational quantum eigensolver},\ }\href {https://doi.org/10.1103/PhysRevResearch.5.013095} {\bibfield  {journal} {\bibinfo  {journal} {Phys. Rev. Res.}\ }\textbf {\bibinfo {volume} {5}},\ \bibinfo {pages} {013095} (\bibinfo {year} {2023})}\BibitemShut {NoStop}%
\bibitem [{\citenamefont {Weaving}\ \emph {et~al.}(2023{\natexlab{b}})\citenamefont {Weaving}, \citenamefont {Ralli}, \citenamefont {Kirby}, \citenamefont {Love}, \citenamefont {Succi},\ and\ \citenamefont {Coveney}}]{weaving2023benchmarking}%
  \BibitemOpen
  \bibfield  {author} {\bibinfo {author} {\bibfnamefont {T.}~\bibnamefont {Weaving}}, \bibinfo {author} {\bibfnamefont {A.}~\bibnamefont {Ralli}}, \bibinfo {author} {\bibfnamefont {W.~M.}\ \bibnamefont {Kirby}}, \bibinfo {author} {\bibfnamefont {P.~J.}\ \bibnamefont {Love}}, \bibinfo {author} {\bibfnamefont {S.}~\bibnamefont {Succi}},\ and\ \bibinfo {author} {\bibfnamefont {P.~V.}\ \bibnamefont {Coveney}},\ }\bibfield  {title} {\bibinfo {title} {Benchmarking noisy intermediate scale quantum error mitigation strategies for ground state preparation of the hcl molecule},\ }\href {https://doi.org/10.1103/PhysRevResearch.5.043054} {\bibfield  {journal} {\bibinfo  {journal} {Phys. Rev. Res.}\ }\textbf {\bibinfo {volume} {5}},\ \bibinfo {pages} {43054} (\bibinfo {year} {2023}{\natexlab{b}})}\BibitemShut {NoStop}%
\bibitem [{\citenamefont {Weaving}\ \emph {et~al.}(2025{\natexlab{a}})\citenamefont {Weaving}, \citenamefont {Ralli}, \citenamefont {Love}, \citenamefont {Succi},\ and\ \citenamefont {Coveney}}]{weaving2023contextual}%
  \BibitemOpen
  \bibfield  {author} {\bibinfo {author} {\bibfnamefont {T.}~\bibnamefont {Weaving}}, \bibinfo {author} {\bibfnamefont {A.}~\bibnamefont {Ralli}}, \bibinfo {author} {\bibfnamefont {P.~J.}\ \bibnamefont {Love}}, \bibinfo {author} {\bibfnamefont {S.}~\bibnamefont {Succi}},\ and\ \bibinfo {author} {\bibfnamefont {P.~V.}\ \bibnamefont {Coveney}},\ }\bibfield  {title} {\bibinfo {title} {Contextual subspace variational quantum eigensolver calculation of the dissociation curve of molecular nitrogen on a superconducting quantum computer},\ }\href {https://doi.org/10.1038/s41534-025-01019-8} {\bibfield  {journal} {\bibinfo  {journal} {npj Quantum Information}\ }\textbf {\bibinfo {volume} {11}},\ \bibinfo {pages} {25} (\bibinfo {year} {2025}{\natexlab{a}})}\BibitemShut {NoStop}%
\bibitem [{\citenamefont {Liang}\ \emph {et~al.}(2023)\citenamefont {Liang}, \citenamefont {Song}, \citenamefont {Cheng}, \citenamefont {Ren}, \citenamefont {Hao}, \citenamefont {Yang}, \citenamefont {Shi},\ and\ \citenamefont {Li}}]{liang2023spacepulse}%
  \BibitemOpen
  \bibfield  {author} {\bibinfo {author} {\bibfnamefont {Z.}~\bibnamefont {Liang}}, \bibinfo {author} {\bibfnamefont {Z.}~\bibnamefont {Song}}, \bibinfo {author} {\bibfnamefont {J.}~\bibnamefont {Cheng}}, \bibinfo {author} {\bibfnamefont {H.}~\bibnamefont {Ren}}, \bibinfo {author} {\bibfnamefont {T.}~\bibnamefont {Hao}}, \bibinfo {author} {\bibfnamefont {R.}~\bibnamefont {Yang}}, \bibinfo {author} {\bibfnamefont {Y.}~\bibnamefont {Shi}},\ and\ \bibinfo {author} {\bibfnamefont {T.}~\bibnamefont {Li}},\ }\bibfield  {title} {\bibinfo {title} {Spacepulse: Combining parameterized pulses and contextual subspace for more practical vqe},\ }\href@noop {} {\bibfield  {journal} {\bibinfo  {journal} {arXiv preprint}\ } (\bibinfo {year} {2023})},\ \Eprint {https://arxiv.org/abs/2311.17423} {2311.17423} \BibitemShut {NoStop}%
\bibitem [{\citenamefont {Kiser}\ \emph {et~al.}(2025)\citenamefont {Kiser}, \citenamefont {Beuerle},\ and\ \citenamefont {Šimkovic IV}}]{kiser2025contextual}%
  \BibitemOpen
  \bibfield  {author} {\bibinfo {author} {\bibfnamefont {M.}~\bibnamefont {Kiser}}, \bibinfo {author} {\bibfnamefont {M.}~\bibnamefont {Beuerle}},\ and\ \bibinfo {author} {\bibfnamefont {F.}~\bibnamefont {Šimkovic IV}},\ }\bibfield  {title} {\bibinfo {title} {Contextual subspace auxiliary-field quantum monte carlo: Improved bias with reduced quantum resources},\ }\href {https://doi.org/10.1021/acs.jctc.4c01280} {\bibfield  {journal} {\bibinfo  {journal} {Journal of Chemical Theory and Computation}\ }\textbf {\bibinfo {volume} {21}},\ \bibinfo {pages} {2256} (\bibinfo {year} {2025})}\BibitemShut {NoStop}%
\bibitem [{\citenamefont {Yao}\ and\ \citenamefont {Li}(2025)}]{yao2025quantum}%
  \BibitemOpen
  \bibfield  {author} {\bibinfo {author} {\bibfnamefont {Q.}~\bibnamefont {Yao}}\ and\ \bibinfo {author} {\bibfnamefont {H.}~\bibnamefont {Li}},\ }\bibfield  {title} {\bibinfo {title} {Quantum inspired excited states calculations for molecules based on contextual subspace and symmetry optimizations},\ }\href@noop {} {\bibfield  {journal} {\bibinfo  {journal} {arXiv preprint}\ } (\bibinfo {year} {2025})},\ \Eprint {https://arxiv.org/abs/2502.17932} {2502.17932} \BibitemShut {NoStop}%
\bibitem [{\citenamefont {Bickley}\ \emph {et~al.}(2025)\citenamefont {Bickley}, \citenamefont {Mingare}, \citenamefont {Weaving}, \citenamefont {de~la Bastida}, \citenamefont {Wan}, \citenamefont {Nibbi}, \citenamefont {Seitz}, \citenamefont {Ralli}, \citenamefont {Love}, \citenamefont {Chung} \emph {et~al.}}]{bickley2025extending}%
  \BibitemOpen
  \bibfield  {author} {\bibinfo {author} {\bibfnamefont {T.~M.}\ \bibnamefont {Bickley}}, \bibinfo {author} {\bibfnamefont {A.}~\bibnamefont {Mingare}}, \bibinfo {author} {\bibfnamefont {T.}~\bibnamefont {Weaving}}, \bibinfo {author} {\bibfnamefont {M.~W.}\ \bibnamefont {de~la Bastida}}, \bibinfo {author} {\bibfnamefont {S.}~\bibnamefont {Wan}}, \bibinfo {author} {\bibfnamefont {M.}~\bibnamefont {Nibbi}}, \bibinfo {author} {\bibfnamefont {P.}~\bibnamefont {Seitz}}, \bibinfo {author} {\bibfnamefont {A.}~\bibnamefont {Ralli}}, \bibinfo {author} {\bibfnamefont {P.~J.}\ \bibnamefont {Love}}, \bibinfo {author} {\bibfnamefont {M.}~\bibnamefont {Chung}}, \emph {et~al.},\ }\bibfield  {title} {\bibinfo {title} {Extending quantum computing through subspace, embedding and classical molecular dynamics techniques},\ }\href@noop {} {\bibfield  {journal} {\bibinfo  {journal} {arXiv preprint}\ } (\bibinfo {year} {2025})},\ \Eprint {https://arxiv.org/abs/2505.16796} {2505.16796} \BibitemShut {NoStop}%
\bibitem [{\citenamefont {Weaving}\ \emph {et~al.}(2025{\natexlab{b}})\citenamefont {Weaving}, \citenamefont {Ralli}, \citenamefont {Love}, \citenamefont {Succi},\ and\ \citenamefont {Coveney}}]{weaving2025accurately}%
  \BibitemOpen
  \bibfield  {author} {\bibinfo {author} {\bibfnamefont {T.}~\bibnamefont {Weaving}}, \bibinfo {author} {\bibfnamefont {A.}~\bibnamefont {Ralli}}, \bibinfo {author} {\bibfnamefont {P.~J.}\ \bibnamefont {Love}}, \bibinfo {author} {\bibfnamefont {S.}~\bibnamefont {Succi}},\ and\ \bibinfo {author} {\bibfnamefont {P.~V.}\ \bibnamefont {Coveney}},\ }\bibfield  {title} {\bibinfo {title} {Accurately simulating the time evolution of an ising model with echo verified clifford data regression on a superconducting quantum computer},\ }\href {https://doi.org/10.22331/q-2025-05-05-1732} {\bibfield  {journal} {\bibinfo  {journal} {Quantum}\ }\textbf {\bibinfo {volume} {9}},\ \bibinfo {pages} {1732} (\bibinfo {year} {2025}{\natexlab{b}})}\BibitemShut {NoStop}%
\bibitem [{\citenamefont {Bravyi}\ \emph {et~al.}(2017)\citenamefont {Bravyi}, \citenamefont {Gambetta}, \citenamefont {Mezzacapo},\ and\ \citenamefont {Temme}}]{bravyi2017tapering}%
  \BibitemOpen
  \bibfield  {author} {\bibinfo {author} {\bibfnamefont {S.}~\bibnamefont {Bravyi}}, \bibinfo {author} {\bibfnamefont {J.~M.}\ \bibnamefont {Gambetta}}, \bibinfo {author} {\bibfnamefont {A.}~\bibnamefont {Mezzacapo}},\ and\ \bibinfo {author} {\bibfnamefont {K.}~\bibnamefont {Temme}},\ }\bibfield  {title} {\bibinfo {title} {Tapering off qubits to simulate fermionic {H}amiltonians},\ }\href@noop {} {\bibfield  {journal} {\bibinfo  {journal} {arXiv preprint}\ } (\bibinfo {year} {2017})},\ \Eprint {https://arxiv.org/abs/arXiv:1701.08213} {arXiv:1701.08213} \BibitemShut {NoStop}%
\bibitem [{\citenamefont {Setia}\ \emph {et~al.}(2020)\citenamefont {Setia}, \citenamefont {Chen}, \citenamefont {Rice}, \citenamefont {Mezzacapo}, \citenamefont {Pistoia},\ and\ \citenamefont {Whitfield}}]{setia2020reducing}%
  \BibitemOpen
  \bibfield  {author} {\bibinfo {author} {\bibfnamefont {K.}~\bibnamefont {Setia}}, \bibinfo {author} {\bibfnamefont {R.}~\bibnamefont {Chen}}, \bibinfo {author} {\bibfnamefont {J.~E.}\ \bibnamefont {Rice}}, \bibinfo {author} {\bibfnamefont {A.}~\bibnamefont {Mezzacapo}}, \bibinfo {author} {\bibfnamefont {M.}~\bibnamefont {Pistoia}},\ and\ \bibinfo {author} {\bibfnamefont {J.~D.}\ \bibnamefont {Whitfield}},\ }\bibfield  {title} {\bibinfo {title} {Reducing qubit requirements for quantum simulations using molecular point group symmetries},\ }\href {https://doi.org/10.1021/acs.jctc.0c00113} {\bibfield  {journal} {\bibinfo  {journal} {Journal of Chemical Theory and Computation}\ }\textbf {\bibinfo {volume} {16}},\ \bibinfo {pages} {6091} (\bibinfo {year} {2020})}\BibitemShut {NoStop}%
\bibitem [{\citenamefont {Gottesman}(1997)}]{gottesman1997stabilizer}%
  \BibitemOpen
  \bibfield  {author} {\bibinfo {author} {\bibfnamefont {D.}~\bibnamefont {Gottesman}},\ }\href@noop {} {\emph {\bibinfo {title} {Stabilizer codes and quantum error correction}}}\ (\bibinfo  {publisher} {California Institute of Technology},\ \bibinfo {year} {1997})\ \Eprint {https://arxiv.org/abs/quant-ph/9705052} {quant-ph/9705052} \BibitemShut {NoStop}%
\bibitem [{\citenamefont {Peres}(1990)}]{peres1990incompatible}%
  \BibitemOpen
  \bibfield  {author} {\bibinfo {author} {\bibfnamefont {A.}~\bibnamefont {Peres}},\ }\bibfield  {title} {\bibinfo {title} {Incompatible results of quantum measurements},\ }\href {https://doi.org/10.1016/0375-9601(90)90172-K} {\bibfield  {journal} {\bibinfo  {journal} {Physics Letters A}\ }\textbf {\bibinfo {volume} {151}},\ \bibinfo {pages} {107} (\bibinfo {year} {1990})}\BibitemShut {NoStop}%
\bibitem [{\citenamefont {Mermin}(1990)}]{mermin1990simple}%
  \BibitemOpen
  \bibfield  {author} {\bibinfo {author} {\bibfnamefont {N.~D.}\ \bibnamefont {Mermin}},\ }\bibfield  {title} {\bibinfo {title} {Simple unified form for the major no-hidden-variables theorems},\ }\href {https://doi.org/10.1103/PhysRevLett.65.3373} {\bibfield  {journal} {\bibinfo  {journal} {Physical review letters}\ }\textbf {\bibinfo {volume} {65}},\ \bibinfo {pages} {3373} (\bibinfo {year} {1990})}\BibitemShut {NoStop}%
\bibitem [{\citenamefont {Mermin}(1993)}]{mermin1993hidden}%
  \BibitemOpen
  \bibfield  {author} {\bibinfo {author} {\bibfnamefont {N.~D.}\ \bibnamefont {Mermin}},\ }\bibfield  {title} {\bibinfo {title} {Hidden variables and the two theorems of john bell},\ }\href {https://doi.org/10.1103/RevModPhys.65.803} {\bibfield  {journal} {\bibinfo  {journal} {Reviews of Modern Physics}\ }\textbf {\bibinfo {volume} {65}},\ \bibinfo {pages} {803} (\bibinfo {year} {1993})}\BibitemShut {NoStop}%
\bibitem [{\citenamefont {Spekkens}(2005)}]{spekkens2005contextuality}%
  \BibitemOpen
  \bibfield  {author} {\bibinfo {author} {\bibfnamefont {R.~W.}\ \bibnamefont {Spekkens}},\ }\bibfield  {title} {\bibinfo {title} {Contextuality for preparations, transformations, and unsharp measurements},\ }\href {https://doi.org/10.1103/PhysRevA.71.052108} {\bibfield  {journal} {\bibinfo  {journal} {Physical Review A—Atomic, Molecular, and Optical Physics}\ }\textbf {\bibinfo {volume} {71}},\ \bibinfo {pages} {052108} (\bibinfo {year} {2005})}\BibitemShut {NoStop}%
\bibitem [{\citenamefont {Spekkens}(2007)}]{spekkens2007evidence}%
  \BibitemOpen
  \bibfield  {author} {\bibinfo {author} {\bibfnamefont {R.~W.}\ \bibnamefont {Spekkens}},\ }\bibfield  {title} {\bibinfo {title} {Evidence for the epistemic view of quantum states: A toy theory},\ }\href {https://doi.org/10.1103/PhysRevA.75.032110} {\bibfield  {journal} {\bibinfo  {journal} {Physical Review A}\ }\textbf {\bibinfo {volume} {75}},\ \bibinfo {pages} {032110} (\bibinfo {year} {2007})}\BibitemShut {NoStop}%
\bibitem [{\citenamefont {Spekkens}(2008)}]{spekkens2008negativity}%
  \BibitemOpen
  \bibfield  {author} {\bibinfo {author} {\bibfnamefont {R.~W.}\ \bibnamefont {Spekkens}},\ }\bibfield  {title} {\bibinfo {title} {Negativity and contextuality are equivalent notions of nonclassicality},\ }\href {https://doi.org/10.1103/PhysRevLett.101.020401} {\bibfield  {journal} {\bibinfo  {journal} {Physical review letters}\ }\textbf {\bibinfo {volume} {101}},\ \bibinfo {pages} {020401} (\bibinfo {year} {2008})}\BibitemShut {NoStop}%
\bibitem [{\citenamefont {Spekkens}(2016)}]{Spekkens2016}%
  \BibitemOpen
  \bibfield  {author} {\bibinfo {author} {\bibfnamefont {R.~W.}\ \bibnamefont {Spekkens}},\ }\bibinfo {title} {Quasi-quantization: Classical statistical theories with an epistemic restriction},\ in\ \href {https://doi.org/10.1007/978-94-017-7303-4_4} {\emph {\bibinfo {booktitle} {Quantum Theory: Informational Foundations and Foils}}}\ (\bibinfo  {publisher} {Springer Netherlands},\ \bibinfo {address} {Dordrecht},\ \bibinfo {year} {2016})\ pp.\ \bibinfo {pages} {83--135}\BibitemShut {NoStop}%
\bibitem [{\citenamefont {Kirby}\ and\ \citenamefont {Love}(2019)}]{kirby2019contextuality}%
  \BibitemOpen
  \bibfield  {author} {\bibinfo {author} {\bibfnamefont {W.~M.}\ \bibnamefont {Kirby}}\ and\ \bibinfo {author} {\bibfnamefont {P.~J.}\ \bibnamefont {Love}},\ }\bibfield  {title} {\bibinfo {title} {Contextuality test of the nonclassicality of variational quantum eigensolvers},\ }\href {https://doi.org/10.1103/PhysRevLett.123.200501} {\bibfield  {journal} {\bibinfo  {journal} {Physical review letters}\ }\textbf {\bibinfo {volume} {123}},\ \bibinfo {pages} {200501} (\bibinfo {year} {2019})}\BibitemShut {NoStop}%
\bibitem [{\citenamefont {Kirby}\ and\ \citenamefont {Love}(2020)}]{kirby2020classical}%
  \BibitemOpen
  \bibfield  {author} {\bibinfo {author} {\bibfnamefont {W.~M.}\ \bibnamefont {Kirby}}\ and\ \bibinfo {author} {\bibfnamefont {P.~J.}\ \bibnamefont {Love}},\ }\bibfield  {title} {\bibinfo {title} {Classical simulation of noncontextual pauli {H}amiltonians},\ }\href {https://doi.org/10.1103/PhysRevA.102.032418} {\bibfield  {journal} {\bibinfo  {journal} {Physical Review A}\ }\textbf {\bibinfo {volume} {102}},\ \bibinfo {pages} {032418} (\bibinfo {year} {2020})}\BibitemShut {NoStop}%
\bibitem [{\citenamefont {Raussendorf}\ \emph {et~al.}(2020)\citenamefont {Raussendorf}, \citenamefont {Bermejo-Vega}, \citenamefont {Tyhurst}, \citenamefont {Okay},\ and\ \citenamefont {Zurel}}]{raussendorf2020phase}%
  \BibitemOpen
  \bibfield  {author} {\bibinfo {author} {\bibfnamefont {R.}~\bibnamefont {Raussendorf}}, \bibinfo {author} {\bibfnamefont {J.}~\bibnamefont {Bermejo-Vega}}, \bibinfo {author} {\bibfnamefont {E.}~\bibnamefont {Tyhurst}}, \bibinfo {author} {\bibfnamefont {C.}~\bibnamefont {Okay}},\ and\ \bibinfo {author} {\bibfnamefont {M.}~\bibnamefont {Zurel}},\ }\bibfield  {title} {\bibinfo {title} {Phase-space-simulation method for quantum computation with magic states on qubits},\ }\href {https://doi.org/10.1103/PhysRevA.101.012350} {\bibfield  {journal} {\bibinfo  {journal} {Physical Review A}\ }\textbf {\bibinfo {volume} {101}},\ \bibinfo {pages} {012350} (\bibinfo {year} {2020})}\BibitemShut {NoStop}%
\bibitem [{\citenamefont {Ralli}\ and\ \citenamefont {Weaving}(2022)}]{symmer2022}%
  \BibitemOpen
  \bibfield  {author} {\bibinfo {author} {\bibfnamefont {A.}~\bibnamefont {Ralli}}\ and\ \bibinfo {author} {\bibfnamefont {T.}~\bibnamefont {Weaving}},\ }\href@noop {} {\bibinfo {title} {symmer}},\ \bibinfo {howpublished} {\url{https://github.com/UCL-CCS/symmer}} (\bibinfo {year} {2022})\BibitemShut {NoStop}%
\bibitem [{\citenamefont {Gray}(2018)}]{Gray2018}%
  \BibitemOpen
  \bibfield  {author} {\bibinfo {author} {\bibfnamefont {J.}~\bibnamefont {Gray}},\ }\bibfield  {title} {\bibinfo {title} {quimb: a python library for quantum information and many-body calculations},\ }\href {https://doi.org/10.21105/joss.00819} {\bibfield  {journal} {\bibinfo  {journal} {Journal of Open Source Software}\ }\textbf {\bibinfo {volume} {3}},\ \bibinfo {pages} {819} (\bibinfo {year} {2018})}\BibitemShut {NoStop}%
\bibitem [{\citenamefont {Izmaylov}\ \emph {et~al.}(2019)\citenamefont {Izmaylov}, \citenamefont {Yen}, \citenamefont {Lang},\ and\ \citenamefont {Verteletskyi}}]{izmaylov2019unitary}%
  \BibitemOpen
  \bibfield  {author} {\bibinfo {author} {\bibfnamefont {A.~F.}\ \bibnamefont {Izmaylov}}, \bibinfo {author} {\bibfnamefont {T.-C.}\ \bibnamefont {Yen}}, \bibinfo {author} {\bibfnamefont {R.~A.}\ \bibnamefont {Lang}},\ and\ \bibinfo {author} {\bibfnamefont {V.}~\bibnamefont {Verteletskyi}},\ }\bibfield  {title} {\bibinfo {title} {Unitary partitioning approach to the measurement problem in the variational quantum eigensolver method},\ }\href {https://doi.org/10.1021/acs.jctc.9b00791} {\bibfield  {journal} {\bibinfo  {journal} {Journal of chemical theory and computation}\ }\textbf {\bibinfo {volume} {16}},\ \bibinfo {pages} {190} (\bibinfo {year} {2019})}\BibitemShut {NoStop}%
\bibitem [{\citenamefont {ANIS}\ \emph {et~al.}(2021)\citenamefont {ANIS} \emph {et~al.}}]{Qiskit}%
  \BibitemOpen
  \bibfield  {author} {\bibinfo {author} {\bibfnamefont {M.~S.}\ \bibnamefont {ANIS}} \emph {et~al.},\ }\href {https://doi.org/10.5281/zenodo.2573505} {\bibinfo {title} {Qiskit: An open-source framework for quantum computing}} (\bibinfo {year} {2021})\BibitemShut {NoStop}%
\bibitem [{\citenamefont {Verteletskyi}\ \emph {et~al.}(2020)\citenamefont {Verteletskyi}, \citenamefont {Yen},\ and\ \citenamefont {Izmaylov}}]{verteletskyi2020measurement}%
  \BibitemOpen
  \bibfield  {author} {\bibinfo {author} {\bibfnamefont {V.}~\bibnamefont {Verteletskyi}}, \bibinfo {author} {\bibfnamefont {T.-C.}\ \bibnamefont {Yen}},\ and\ \bibinfo {author} {\bibfnamefont {A.~F.}\ \bibnamefont {Izmaylov}},\ }\bibfield  {title} {\bibinfo {title} {Measurement optimization in the variational quantum eigensolver using a minimum clique cover},\ }\href {https://doi.org/10.1063/1.5141458} {\bibfield  {journal} {\bibinfo  {journal} {The Journal of chemical physics}\ }\textbf {\bibinfo {volume} {152}},\ \bibinfo {pages} {124114} (\bibinfo {year} {2020})}\BibitemShut {NoStop}%
\bibitem [{\citenamefont {Yen}\ \emph {et~al.}(2020)\citenamefont {Yen}, \citenamefont {Verteletskyi},\ and\ \citenamefont {Izmaylov}}]{yen2020measuring}%
  \BibitemOpen
  \bibfield  {author} {\bibinfo {author} {\bibfnamefont {T.-C.}\ \bibnamefont {Yen}}, \bibinfo {author} {\bibfnamefont {V.}~\bibnamefont {Verteletskyi}},\ and\ \bibinfo {author} {\bibfnamefont {A.~F.}\ \bibnamefont {Izmaylov}},\ }\bibfield  {title} {\bibinfo {title} {Measuring all compatible operators in one series of single-qubit measurements using unitary transformations},\ }\href {https://doi.org/10.1021/acs.jctc.0c00008} {\bibfield  {journal} {\bibinfo  {journal} {Journal of chemical theory and computation}\ }\textbf {\bibinfo {volume} {16}},\ \bibinfo {pages} {2400} (\bibinfo {year} {2020})}\BibitemShut {NoStop}%
\bibitem [{\citenamefont {Zhao}\ \emph {et~al.}(2020)\citenamefont {Zhao}, \citenamefont {Tranter}, \citenamefont {Kirby}, \citenamefont {Ung}, \citenamefont {Miyake},\ and\ \citenamefont {Love}}]{zhao2020measurement}%
  \BibitemOpen
  \bibfield  {author} {\bibinfo {author} {\bibfnamefont {A.}~\bibnamefont {Zhao}}, \bibinfo {author} {\bibfnamefont {A.}~\bibnamefont {Tranter}}, \bibinfo {author} {\bibfnamefont {W.~M.}\ \bibnamefont {Kirby}}, \bibinfo {author} {\bibfnamefont {S.~F.}\ \bibnamefont {Ung}}, \bibinfo {author} {\bibfnamefont {A.}~\bibnamefont {Miyake}},\ and\ \bibinfo {author} {\bibfnamefont {P.~J.}\ \bibnamefont {Love}},\ }\bibfield  {title} {\bibinfo {title} {Measurement reduction in variational quantum algorithms},\ }\href {https://doi.org/10.1103/PhysRevA.101.062322} {\bibfield  {journal} {\bibinfo  {journal} {Physical Review A}\ }\textbf {\bibinfo {volume} {101}},\ \bibinfo {pages} {062322} (\bibinfo {year} {2020})}\BibitemShut {NoStop}%
\bibitem [{\citenamefont {Fletcher}(2013)}]{fletcher2013practical}%
  \BibitemOpen
  \bibfield  {author} {\bibinfo {author} {\bibfnamefont {R.}~\bibnamefont {Fletcher}},\ }\href {https://doi.org/10.1002/9781118723203} {\emph {\bibinfo {title} {Practical methods of optimization}}}\ (\bibinfo  {publisher} {John Wiley \& Sons},\ \bibinfo {year} {2013})\BibitemShut {NoStop}%
\bibitem [{\citenamefont {Hestenes}\ \emph {et~al.}(1952)\citenamefont {Hestenes}, \citenamefont {Stiefel} \emph {et~al.}}]{hestenes1952methods}%
  \BibitemOpen
  \bibfield  {author} {\bibinfo {author} {\bibfnamefont {M.~R.}\ \bibnamefont {Hestenes}}, \bibinfo {author} {\bibfnamefont {E.}~\bibnamefont {Stiefel}}, \emph {et~al.},\ }\href {https://doi.org/10.6028/jres.049.044} {\emph {\bibinfo {title} {Methods of conjugate gradients for solving linear systems}}},\ Vol.~\bibinfo {volume} {49}\ (\bibinfo  {publisher} {NBS Washington, DC},\ \bibinfo {year} {1952})\BibitemShut {NoStop}%
\bibitem [{\citenamefont {Virtanen}\ \emph {et~al.}(2020)\citenamefont {Virtanen} \emph {et~al.}}]{2020SciPy-NMeth}%
  \BibitemOpen
  \bibfield  {author} {\bibinfo {author} {\bibfnamefont {P.}~\bibnamefont {Virtanen}} \emph {et~al.},\ }\bibfield  {title} {\bibinfo {title} {{{SciPy} 1.0: Fundamental Algorithms for Scientific Computing in Python}},\ }\href {https://doi.org/10.1038/s41592-019-0686-2} {\bibfield  {journal} {\bibinfo  {journal} {Nature Methods}\ }\textbf {\bibinfo {volume} {17}},\ \bibinfo {pages} {261} (\bibinfo {year} {2020})}\BibitemShut {NoStop}%
\bibitem [{\citenamefont {Parrish}\ \emph {et~al.}(2019)\citenamefont {Parrish}, \citenamefont {Hohenstein}, \citenamefont {McMahon},\ and\ \citenamefont {Martinez}}]{parrish2019hybrid}%
  \BibitemOpen
  \bibfield  {author} {\bibinfo {author} {\bibfnamefont {R.~M.}\ \bibnamefont {Parrish}}, \bibinfo {author} {\bibfnamefont {E.~G.}\ \bibnamefont {Hohenstein}}, \bibinfo {author} {\bibfnamefont {P.~L.}\ \bibnamefont {McMahon}},\ and\ \bibinfo {author} {\bibfnamefont {T.~J.}\ \bibnamefont {Martinez}},\ }\bibfield  {title} {\bibinfo {title} {Hybrid quantum/classical derivative theory: Analytical gradients and excited-state dynamics for the multistate contracted variational quantum eigensolver},\ }\href@noop {} {\bibfield  {journal} {\bibinfo  {journal} {arXiv preprint}\ } (\bibinfo {year} {2019})},\ \Eprint {https://arxiv.org/abs/1906.08728} {1906.08728} \BibitemShut {NoStop}%
\bibitem [{\citenamefont {Nation}\ \emph {et~al.}(2021)\citenamefont {Nation}, \citenamefont {Kang}, \citenamefont {Sundaresan},\ and\ \citenamefont {Gambetta}}]{nation2021scalable}%
  \BibitemOpen
  \bibfield  {author} {\bibinfo {author} {\bibfnamefont {P.~D.}\ \bibnamefont {Nation}}, \bibinfo {author} {\bibfnamefont {H.}~\bibnamefont {Kang}}, \bibinfo {author} {\bibfnamefont {N.}~\bibnamefont {Sundaresan}},\ and\ \bibinfo {author} {\bibfnamefont {J.~M.}\ \bibnamefont {Gambetta}},\ }\bibfield  {title} {\bibinfo {title} {Scalable mitigation of measurement errors on quantum computers},\ }\href {https://doi.org/10.1103/PRXQuantum.2.040326} {\bibfield  {journal} {\bibinfo  {journal} {PRX Quantum}\ }\textbf {\bibinfo {volume} {2}},\ \bibinfo {pages} {040326} (\bibinfo {year} {2021})}\BibitemShut {NoStop}%
\bibitem [{\citenamefont {Bravyi}\ \emph {et~al.}(2021)\citenamefont {Bravyi}, \citenamefont {Sheldon}, \citenamefont {Kandala}, \citenamefont {Mckay},\ and\ \citenamefont {Gambetta}}]{bravyi2021mitigating}%
  \BibitemOpen
  \bibfield  {author} {\bibinfo {author} {\bibfnamefont {S.}~\bibnamefont {Bravyi}}, \bibinfo {author} {\bibfnamefont {S.}~\bibnamefont {Sheldon}}, \bibinfo {author} {\bibfnamefont {A.}~\bibnamefont {Kandala}}, \bibinfo {author} {\bibfnamefont {D.~C.}\ \bibnamefont {Mckay}},\ and\ \bibinfo {author} {\bibfnamefont {J.~M.}\ \bibnamefont {Gambetta}},\ }\bibfield  {title} {\bibinfo {title} {Mitigating measurement errors in multiqubit experiments},\ }\href {https://doi.org/10.1103/PhysRevA.103.042605} {\bibfield  {journal} {\bibinfo  {journal} {Physical Review A}\ }\textbf {\bibinfo {volume} {103}},\ \bibinfo {pages} {042605} (\bibinfo {year} {2021})}\BibitemShut {NoStop}%
\bibitem [{\citenamefont {van~den Berg}\ \emph {et~al.}(2022)\citenamefont {van~den Berg}, \citenamefont {Minev},\ and\ \citenamefont {Temme}}]{van2022model}%
  \BibitemOpen
  \bibfield  {author} {\bibinfo {author} {\bibfnamefont {E.}~\bibnamefont {van~den Berg}}, \bibinfo {author} {\bibfnamefont {Z.~K.}\ \bibnamefont {Minev}},\ and\ \bibinfo {author} {\bibfnamefont {K.}~\bibnamefont {Temme}},\ }\bibfield  {title} {\bibinfo {title} {Model-free readout-error mitigation for quantum expectation values},\ }\href {https://doi.org/10.1103/PhysRevA.105.032620} {\bibfield  {journal} {\bibinfo  {journal} {Physical Review A}\ }\textbf {\bibinfo {volume} {105}},\ \bibinfo {pages} {032620} (\bibinfo {year} {2022})}\BibitemShut {NoStop}%
\bibitem [{\citenamefont {Bonet-Monroig}\ \emph {et~al.}(2018)\citenamefont {Bonet-Monroig}, \citenamefont {Sagastizabal}, \citenamefont {Singh},\ and\ \citenamefont {O'Brien}}]{bonet2018low}%
  \BibitemOpen
  \bibfield  {author} {\bibinfo {author} {\bibfnamefont {X.}~\bibnamefont {Bonet-Monroig}}, \bibinfo {author} {\bibfnamefont {R.}~\bibnamefont {Sagastizabal}}, \bibinfo {author} {\bibfnamefont {M.}~\bibnamefont {Singh}},\ and\ \bibinfo {author} {\bibfnamefont {T.}~\bibnamefont {O'Brien}},\ }\bibfield  {title} {\bibinfo {title} {Low-cost error mitigation by symmetry verification},\ }\href {https://doi.org/10.1103/PhysRevA.98.062339} {\bibfield  {journal} {\bibinfo  {journal} {Physical Review A}\ }\textbf {\bibinfo {volume} {98}},\ \bibinfo {pages} {062339} (\bibinfo {year} {2018})}\BibitemShut {NoStop}%
\bibitem [{\citenamefont {McArdle}\ \emph {et~al.}(2019)\citenamefont {McArdle}, \citenamefont {Yuan},\ and\ \citenamefont {Benjamin}}]{mcardle2019error}%
  \BibitemOpen
  \bibfield  {author} {\bibinfo {author} {\bibfnamefont {S.}~\bibnamefont {McArdle}}, \bibinfo {author} {\bibfnamefont {X.}~\bibnamefont {Yuan}},\ and\ \bibinfo {author} {\bibfnamefont {S.}~\bibnamefont {Benjamin}},\ }\bibfield  {title} {\bibinfo {title} {Error-mitigated digital quantum simulation},\ }\href {https://doi.org/10.1103/PhysRevLett.122.180501} {\bibfield  {journal} {\bibinfo  {journal} {Physical review letters}\ }\textbf {\bibinfo {volume} {122}},\ \bibinfo {pages} {180501} (\bibinfo {year} {2019})}\BibitemShut {NoStop}%
\bibitem [{\citenamefont {Cai}(2021)}]{cai2021quantum}%
  \BibitemOpen
  \bibfield  {author} {\bibinfo {author} {\bibfnamefont {Z.}~\bibnamefont {Cai}},\ }\bibfield  {title} {\bibinfo {title} {Quantum error mitigation using symmetry expansion},\ }\href {https://doi.org/10.22331/q-2021-09-21-548} {\bibfield  {journal} {\bibinfo  {journal} {Quantum}\ }\textbf {\bibinfo {volume} {5}},\ \bibinfo {pages} {548} (\bibinfo {year} {2021})}\BibitemShut {NoStop}%
\bibitem [{\citenamefont {Li}\ and\ \citenamefont {Benjamin}(2017)}]{li2017efficient}%
  \BibitemOpen
  \bibfield  {author} {\bibinfo {author} {\bibfnamefont {Y.}~\bibnamefont {Li}}\ and\ \bibinfo {author} {\bibfnamefont {S.~C.}\ \bibnamefont {Benjamin}},\ }\bibfield  {title} {\bibinfo {title} {Efficient variational quantum simulator incorporating active error minimization},\ }\href {https://doi.org/10.1103/PhysRevX.7.021050} {\bibfield  {journal} {\bibinfo  {journal} {Physical Review X}\ }\textbf {\bibinfo {volume} {7}},\ \bibinfo {pages} {021050} (\bibinfo {year} {2017})}\BibitemShut {NoStop}%
\bibitem [{\citenamefont {Temme}\ \emph {et~al.}(2017)\citenamefont {Temme}, \citenamefont {Bravyi},\ and\ \citenamefont {Gambetta}}]{temme2017error}%
  \BibitemOpen
  \bibfield  {author} {\bibinfo {author} {\bibfnamefont {K.}~\bibnamefont {Temme}}, \bibinfo {author} {\bibfnamefont {S.}~\bibnamefont {Bravyi}},\ and\ \bibinfo {author} {\bibfnamefont {J.~M.}\ \bibnamefont {Gambetta}},\ }\bibfield  {title} {\bibinfo {title} {Error mitigation for short-depth quantum circuits},\ }\href {https://doi.org/10.1103/PhysRevLett.119.180509} {\bibfield  {journal} {\bibinfo  {journal} {Physical review letters}\ }\textbf {\bibinfo {volume} {119}},\ \bibinfo {pages} {180509} (\bibinfo {year} {2017})}\BibitemShut {NoStop}%
\bibitem [{\citenamefont {Endo}\ \emph {et~al.}(2018)\citenamefont {Endo}, \citenamefont {Benjamin},\ and\ \citenamefont {Li}}]{endo2018practical}%
  \BibitemOpen
  \bibfield  {author} {\bibinfo {author} {\bibfnamefont {S.}~\bibnamefont {Endo}}, \bibinfo {author} {\bibfnamefont {S.~C.}\ \bibnamefont {Benjamin}},\ and\ \bibinfo {author} {\bibfnamefont {Y.}~\bibnamefont {Li}},\ }\bibfield  {title} {\bibinfo {title} {Practical quantum error mitigation for near-future applications},\ }\href {https://doi.org/10.1103/PhysRevX.8.031027} {\bibfield  {journal} {\bibinfo  {journal} {Physical Review X}\ }\textbf {\bibinfo {volume} {8}},\ \bibinfo {pages} {031027} (\bibinfo {year} {2018})}\BibitemShut {NoStop}%
\bibitem [{\citenamefont {Kandala}\ \emph {et~al.}(2019)\citenamefont {Kandala}, \citenamefont {Temme}, \citenamefont {C{\'o}rcoles}, \citenamefont {Mezzacapo}, \citenamefont {Chow},\ and\ \citenamefont {Gambetta}}]{kandala2019error}%
  \BibitemOpen
  \bibfield  {author} {\bibinfo {author} {\bibfnamefont {A.}~\bibnamefont {Kandala}}, \bibinfo {author} {\bibfnamefont {K.}~\bibnamefont {Temme}}, \bibinfo {author} {\bibfnamefont {A.~D.}\ \bibnamefont {C{\'o}rcoles}}, \bibinfo {author} {\bibfnamefont {A.}~\bibnamefont {Mezzacapo}}, \bibinfo {author} {\bibfnamefont {J.~M.}\ \bibnamefont {Chow}},\ and\ \bibinfo {author} {\bibfnamefont {J.~M.}\ \bibnamefont {Gambetta}},\ }\bibfield  {title} {\bibinfo {title} {Error mitigation extends the computational reach of a noisy quantum processor},\ }\href {https://doi.org/10.1038/s41586-019-1040-7} {\bibfield  {journal} {\bibinfo  {journal} {Nature}\ }\textbf {\bibinfo {volume} {567}},\ \bibinfo {pages} {491} (\bibinfo {year} {2019})}\BibitemShut {NoStop}%
\bibitem [{\citenamefont {Giurgica-Tiron}\ \emph {et~al.}(2020)\citenamefont {Giurgica-Tiron}, \citenamefont {Hindy}, \citenamefont {LaRose}, \citenamefont {Mari},\ and\ \citenamefont {Zeng}}]{giurgica2020digital}%
  \BibitemOpen
  \bibfield  {author} {\bibinfo {author} {\bibfnamefont {T.}~\bibnamefont {Giurgica-Tiron}}, \bibinfo {author} {\bibfnamefont {Y.}~\bibnamefont {Hindy}}, \bibinfo {author} {\bibfnamefont {R.}~\bibnamefont {LaRose}}, \bibinfo {author} {\bibfnamefont {A.}~\bibnamefont {Mari}},\ and\ \bibinfo {author} {\bibfnamefont {W.~J.}\ \bibnamefont {Zeng}},\ }\bibfield  {title} {\bibinfo {title} {Digital zero noise extrapolation for quantum error mitigation},\ }in\ \href {https://doi.org/10.1109/QCE49297.2020.00045} {\emph {\bibinfo {booktitle} {2020 IEEE International Conference on Quantum Computing and Engineering (QCE)}}}\ (\bibinfo {organization} {IEEE},\ \bibinfo {year} {2020})\ pp.\ \bibinfo {pages} {306--316}\BibitemShut {NoStop}%
\bibitem [{\citenamefont {He}\ \emph {et~al.}(2020)\citenamefont {He}, \citenamefont {Nachman}, \citenamefont {de~Jong},\ and\ \citenamefont {Bauer}}]{he2020zero}%
  \BibitemOpen
  \bibfield  {author} {\bibinfo {author} {\bibfnamefont {A.}~\bibnamefont {He}}, \bibinfo {author} {\bibfnamefont {B.}~\bibnamefont {Nachman}}, \bibinfo {author} {\bibfnamefont {W.~A.}\ \bibnamefont {de~Jong}},\ and\ \bibinfo {author} {\bibfnamefont {C.~W.}\ \bibnamefont {Bauer}},\ }\bibfield  {title} {\bibinfo {title} {Zero-noise extrapolation for quantum-gate error mitigation with identity insertions},\ }\href {https://doi.org/10.1103/PhysRevA.102.012426} {\bibfield  {journal} {\bibinfo  {journal} {Physical Review A}\ }\textbf {\bibinfo {volume} {102}},\ \bibinfo {pages} {012426} (\bibinfo {year} {2020})}\BibitemShut {NoStop}%
\bibitem [{\citenamefont {Mari}\ \emph {et~al.}(2021)\citenamefont {Mari}, \citenamefont {Shammah},\ and\ \citenamefont {Zeng}}]{mari2021extending}%
  \BibitemOpen
  \bibfield  {author} {\bibinfo {author} {\bibfnamefont {A.}~\bibnamefont {Mari}}, \bibinfo {author} {\bibfnamefont {N.}~\bibnamefont {Shammah}},\ and\ \bibinfo {author} {\bibfnamefont {W.~J.}\ \bibnamefont {Zeng}},\ }\bibfield  {title} {\bibinfo {title} {Extending quantum probabilistic error cancellation by noise scaling},\ }\href {https://doi.org/10.1103/PhysRevA.104.052607} {\bibfield  {journal} {\bibinfo  {journal} {Physical Review A}\ }\textbf {\bibinfo {volume} {104}},\ \bibinfo {pages} {052607} (\bibinfo {year} {2021})}\BibitemShut {NoStop}%
\end{thebibliography}%

\end{document}